\documentclass[12pt]{article}
\usepackage{amsmath,amssymb}
\usepackage[hypertex]{hyperref}

\numberwithin{equation}{section}

\def\cN{{\mathcal N}}
\def\cO{{\mathcal O}}

\def\ds{\displaystyle}
\def\ssp{\hspace{0.3mm}}
\def\({\left(}
\def\){\right)}
\renewcommand{\[}{\left[}
\renewcommand{\]}{\right]}
\def\bb#1{\mathbb{#1}}
\def\bmt#1{{\mbox{\boldmath$ #1 $}}}
\def\pare#1{\left\{ #1\right\}}

\def\abs#1{\left| #1\right|}

\def\AdSxS{{AdS${}_5 \times {}$S${}^5$}}
\renewcommand{\eqref}[1]{$\({\rm \ref{#1}}\)$}

\def\Res{\mathop {\rm Res} \limits}

\DeclareMathOperator{\arcsinh}{arcsinh}

\begin{document}

{\ }
\vspace{-10mm}

\begin{flushright}
{\bf July 2008}\\[1mm]
{\small TCDMATH 08\,-\,09}\\[1mm]
{\small UT\,-\,08\,-\,22}
\end{flushright}

\vskip 2cm

\begin{center}
\LARGE

\mbox{\bf Finite\,-\,Size Effects for Multi\,-\,Magnon States}


\vskip 2cm
\renewcommand{\thefootnote}{$\alph{footnote}$}

\large
\centerline{\sc
Yasuyuki Hatsuda$^{\dagger,\,}$\footnote{{\tt \,hatsuda@hep-th.phys.s.u-tokyo.ac.jp}}
\quad and \quad\ 
Ryo Suzuki$^{\ddagger,\,}$\footnote{{\tt \,rsuzuki@maths.tcd.ie}}}

\vskip 1cm

${}^{\dagger}$\emph{Department of Physics, Faculty of Science, University of Tokyo,\\
Bunkyo-ku, Tokyo 113-0033, Japan}

\vspace{0.3cm}

${}^{\ddagger}$\emph{School of Mathematics, Trinity College, Dublin 2, Ireland}

\end{center}

\vskip 14mm

\centerline{\bf Abstract}

\vskip 6mm

We propose the generalized L\"uscher formula for multi-magnon states by which one can compute the finite-size correction to the energy of multi giant magnons at classical and one-loop levels. It is shown that the $F$-term of our formula is consistent with the exact finite-size spectrum of the sinh-Gordon model, and the $\mu$-term agrees with the finite-size energy of magnon boundstate of the asymptotic Bethe Ansatz in the $su(2)$ sector at strong coupling.
In an appendix, we evaluate our formula at weak coupling under some approximations, and find that the transcendental terms arise from a sum over an infinite tower of BPS boundstates.

\vspace*{1.0cm}

\vfill
\thispagestyle{empty}
\setcounter{page}{0}
\setcounter{footnote}{0}
\renewcommand{\thefootnote}{\arabic{footnote}}
\newpage

\section{Introduction}\label{sec:intro}

Years of intensive study of the AdS/CFT correspondence \cite{Maldacena97, GKP98, Witten98} have clarified various aspects of gauge theory and string theory. Most notably, the integrability of $\cN=4$ super Yang-Mills \cite{MZ02, BKS03} and classical superstring on \AdSxS\ \cite{BPR03, KMMZ04, BKSZ05a, BKSZ05b} provided us with a powerful tool to examine the correspondence of the spectrum of both theories. Furthermore, the asymptotic Bethe Ansatz \cite{Staudacher04, BS05} with the dressing phase \cite{AFS04, HL06, Janik06, AF06, BHL06, BES06, BCDKS06, BMR07} succeeded in computing the conformal dimension of super Yang-Mills operators with large $R$-charge at weak coupling, and the energy of closed string states with a large angular momentum at strong coupling, by mapping an operator or a string state to a spin chain state.

The structure of integrability often becomes simplified when we take one of the global charges to a very large value or infinity. In particular, a spin chain with $su(2|2)^2$ symmetry appears when the size, {\it i.e.} the $R$-charge of an operator or the angular momentum of a string, becomes infinite \cite{Beisert05}. The spectrum of the $su(2|2)^2$ spin chain contains an infinite tower of BPS `magnon' boundstates with non-trivial central charges \cite{Dorey06, Beisert06b, CDO06c}. String theory duals of these BPS objects are called giant magnons \cite{HM06} or dyonic giant magnons \cite{CDO06a}. On gauge theory side, these magnons or magnon boundstates are believed to obey the dispersion relation
\begin{equation}
\Delta - J_1 = \sqrt{Q^2 + \frac{\lambda}{\pi^2} \, \sin^2 \frac{p}{2}} \,, \qquad \( \Delta, J_1 \to \infty \),
\label{magnon dispersion:gauge}
\end{equation}
where $\Delta$ is the conformal dimension, $J_1$ is one of the $R$-charges of the super Yang-Mills operator, $Q$ is the number of constituent magnons in a boundstate, $\lambda$ is the 't Hooft coupling and $p$ is the magnon momentum. On string theory side, giant magnons or dyonic giant magnons satisfy
\begin{equation}
E - J_1 = \sqrt{J_2^2 + \frac{\lambda}{\pi^2} \, \sin^2 \frac{\Delta \varphi_1}{2}} \,, \qquad \( E, J_1 \to \infty \),
\label{magnon dispersion:string}
\end{equation}
where $E$ is the energy, $J_1 \,, J_2$ are the angular momenta of a string, and $\Delta \varphi_1$ is the angular difference between string endpoints.
The remarkable agreement between \eqref{magnon dispersion:gauge} and \eqref{magnon dispersion:string} demonstrates that the AdS/CFT correspondence is highly trustable and the integrability will be a promising approach. Details of the related progress including historical perspective can be found in the review article of \cite{Okamura08}.

When $J_1$ is finite, magnons or magnon boundstates receive finite-size corrections to their energy \eqref{magnon dispersion:gauge} or \eqref{magnon dispersion:string}, which behave quite differently at weak coupling and at strong coupling.

At weak coupling, the asymptotic Bethe Ansatz correctly predicts the conformal dimension roughly up to the order of $\lambda^L$ with $L$ the length of the spin chain. Beyond that order, virtual particles start to wrap around the spin chain, modifying the energy accordingly. Finite-size effects of this type are called wrapping effects \cite{BDS04, ST05, AJK05}, and recent calculations showed the anomalous dimension of length-four Konishi operator disagrees with the prediction of the asymptotic Bethe Ansatz \cite{BES06} starting from four-loop in $\lambda$ \cite{KLRSV07, FSSZ07, FSSZ08a, KM08},
\begin{equation}
\begin{array}{rll}
\Delta_{\rm Konishi} &= 4 + 12 g^2 - 48 g^4 + 336 g^6  + \Delta^{(4)} g^8  \,, &\quad \ds g \equiv \frac{\sqrt \lambda}{4 \pi}\,.
\end{array}
\label{Konishi wrapping}
\end{equation}
The prediction of the asymptotic Bethe Ansatz is $\Delta^{(4)}_{\rm ABA} = -2820 - 288 \zeta (3)$. There are two different predictions based on calculation of Feynman diagrams \cite{FSSZ07, FSSZ08a, KM08}, but both of the diagrammatic results show the appearance of new degree 5 of transcendentality which never appeared in the prediction of the asymptotic Bethe Ansatz.\footnote{%
Note added: The very recent work \cite{BJ08} strongly supports that $\Delta^{(4)}$ in \cite{FSSZ07, FSSZ08a} is the correct four-loop anomalous dimension of the length four Konishi operator.
}

At strong coupling, finite-size (or finite-$J_1$) correction to giant magnons or dyonic giant magnons comes in as a term exponentially suppressed in size at classical level \cite{AFZ06a, AFGS07, HS08, MO08, KlMc08} as well as one-loop level \cite{GSV08, HJL08}. Similar behavior was observed earlier in \cite{Sakura06, SZZ06}, where it was found that the quantum string Bethe Ansatz could not reproduce such exponentially suppressed one-loop terms \cite{SZZ06}.

\bigskip
Assuming that the integrability in both gauge and string theories still survives at finite size, we are motivated to develop an appropriate method to compute finite-size effects from infinite-size information. Interestingly, it was argued that wrapping effects at weak coupling are related to exponential corrections at strong coupling \cite{AJK05, RSS05}. Thus we may hope that both effects are explained in a simple and unified way.

Several methods to compute finite-size spectrum are known in the literature of integrable systems. Considering the fact that (giant) magnons are excitations over the ferromagnetic vacuum of the $su(2|2)^2$ spin chain, we think of the following two methods as hopeful candidates: the L\"uscher formula \cite{Luscher83, Luscher86a} (see also \cite{KM91, KK04, KK05}) and the Thermodynamic Bethe Ansatz \cite{Zamolodchikov90, DT96, DT97}.

The original L\"uscher formula relates finite-size mass shifts in relativistic field theories in any dimensions with the $S$-matrix of the infinite-size theory. Conversely said, once the exact finite-size mass is known, through this formula one can probe the spectrum and the $S$-matrix of infinite-size theory in great detail.
The L\"uscher formula was generalized to the case of general dispersion relation by Janik and \L ukowski \cite{JL07}, where they showed it reproduced the leading finite-size correction to a classical giant magnon \cite{AFZ06a, AFGS07}.
There they had to evaluate carefully the dressing phase to all orders in $\lambda$ to obtain the correct answer. This agreement provided another consistency check for the conjectured expression of dressing phase.
Similar agreements are also found for a classical dyonic giant magnon \cite{HS08}, and the results for a one-loop giant magnon are found in \cite{GSV08, HJL08}.

The (generalized) L\"uscher formula is applicable to a large variety of field theories and thus quite useful. However, the formula is valid only when the system size is very large. In contrast, Thermodynamic Bethe Ansatz (TBA) is believed to give the exact spectrum at arbitrary size, although applicable only to integrable systems with factorized scattering. When the system size becomes large, TBA for one-particle states agrees with the (generalized) L\"uscher formula \cite{DT96, DT97}. We expect that in any integrable systems to which Bethe Ansatz is applicable, one can obtain the generalized L\"uscher formula for multi-particle states by taking the large size limit of TBA for the corresponding states.

However, it turns out that the formulation of TBA for the $su(2|2)^2$ spin chain is not at all easy. For this purpose, we need to know the complete spectrum and the $S$-matrix of so-called mirror theory \cite{AF07}. Although it is conjectured the mirror $S$-matrix is related to $S$-matrix of the original theory by analytic continuation \cite{AF07}, explicit computation of the elementary-boundstate and boundstate-boundstate $S$-matrices looks quite complicated \cite{AF08a}.

\bigskip
The aim of this paper is to propose the generalized L\"uscher formula for multi-particle states.
A candidate of such formula has recently been conjectured in \cite{PV08} without justification.
Though we do not derive it from the TBA equations for the $su(2|2)^2$ spin chain, we collect positive supports for our proposal through comparison with semiclassical strings on \AdSxS, the exact finite-size spectrum of sinh-Gordon theory \cite{Teschner07}, and the computation of finite-size correction to the energy of multi magnon states from the Bethe Ansatz \cite{Pozsgay08}.

\bigskip
In an appendix, we apply our formula to two magnon states in the $su(2)$ sector at weak coupling, including the length four Konishi descendant of $\cN=4$ theory. Interestingly, it has recently been indicated that  wrapping effects for such operators exhibits transcendentality depending on the length of spin chain \cite{FSSZ08b}, namely
\begin{equation}
\begin{array}{l}
\text{The wrapping effects of length $L$ operator in the $su(2)$ sector} \\
\text{receives corrections of the form $\zeta (2L-3)$ starting from $g^{2L}$.}
\end{array}
\label{wrap-trans:quote}
\end{equation}
Although this observation is obtained for one-magnon states of the $\beta$-deformed super Yang-Mills, we guess that this general pattern will also hold for (two-magnon) states in $\cN=4$ theory, as it does for $L=4$ \eqref{Konishi wrapping}.
Note that transcendental feature of wrapping effects was also found in \cite{PV08}, in which they considered sum over an infinite species of auxiliary roots in $sl(2)$ spin chain, though the degree of transcendentality was different from \eqref{wrap-trans:quote} for their toy model.

As soon as one tries to apply our formula to length $L$ operators at weak coupling, one finds that just summing over elementary particles does not reproduce transcendental nature of wrapping effects. To solve this problem, we propose to execute summation over an infinite tower of boundstates to reproduce the correct wrapping behavior. This prescription seems quite natural from a TBA point of view, where we obtain the partition function by summing over all particle spectrum of the (mirror) theory.

\bigskip
This paper is organized as follows. In Section \ref{sec:luscher} we briefly review the generalized L\"uscher formulae for one-particle states and present our proposal for multi-particle states.
In Section \ref{sec:F-term} we derive our $F$-term formula by slightly generalizing one-loop finite-gap computation of \cite{GSV08}, and further compare the $F$-term formula with Teschner's exact results for sinh-Gordon model \cite{Teschner07}.
In Section \ref{sec:mu-term} we discuss the $\mu$-term, which reproduces the leading finite-size behavior of multi (dyonic) giant magnons found in \cite{MO08}. For boundstates in the $su(2)$ sector, the $\mu$-term formula can be derived from the asymptotic Bethe Ansatz under some assumptions. In Section \ref{sec:summary} we give summary and discussions.
Some details of computation are explained in Appendices \ref{app:FG} and \ref{sec:notation}. Finally in Appendix \ref{sec:wrapping}, we apply our formula to length $L$ multi-magnon states at weak coupling.

\bigskip \noindent
{\bf Note added:}

While this paper is in preparation, we find a paper \cite{BJ08} on arXiv, which has a substantial overlap with ours.

\section{The generalized L\"uscher formula}\label{sec:luscher}

We start from a brief description of the generalized L\"uscher formula for one-particle states, and then propose the refined formula for multi-particle states.

\subsection{One-particle states}

A particle in quantum field theory is always accompanied by a cloud of virtual particles. When it is put on a periodic box of length $L$, virtual particles start to wrap around, polarizing the vacuum. They also modify the energy of real particles, and this finite-$L$ correction should come as $e^{-qL}$, where $q$ is the Euclidean momentum of the virtual particle. This is the finite-size effect computed by L\"uscher's $F$-term formula. In addition, when virtual particles interact with a real (boundstate) particle, they may induce the decay or the merge of the boundstate. Such effects are described by L\"uscher's $\mu$-term formula.

Let $a$ be an incoming particle with real momentum $p$, and $b$ be a virtual particle with momentum $q$ wrapping around the cylinder. We denote the finite-size correction to the energy of particle $a$ by $\delta E_a$\,. The generalized L\"uscher formula for one-particle state proposed by Janik and \L ukowski is written as $\delta E_a = \delta E_a^F + \delta E_a^\mu$\,. The $F$-term is given by
\begin{equation}
\delta E_a^F = -\sum_b (-1)^{F_b} \int_{-\infty}^\infty \frac{d \tilde q}{2\pi} \(1-\frac{\epsilon_a'(p)}{\epsilon_b'(q^1)}\) e^{-iq^1L} \( S_{ba}^{ba}(q^1,p) - 1 \) \,,
\label{JL F-term}
\end{equation}
where $\tilde q = i q^0$ is the Euclidean energy of the virtual particle, $\epsilon_a (p)$ is the dispersion relation of particle $a$, and $\epsilon_a' (p) = (d \epsilon_a (p)/dp)$. The symbol $F_b$ accounts for the statistics, and takes the value $+1$ if $b$ is a boson and $-1$ if $b$ is a fermion. Here the `virtual' particle $b$ has already been put on-shell because off-shell contribution can be neglected. Thus the particle $b$ obeys the condition $\tilde q^2 + \epsilon_b^2 (q^1) = 0$. The $\mu$-term arises if the integral over $\tilde{q}$ in \eqref{JL F-term} picks up a pole of the $S$-matrix, and is given by
\begin{equation}
\delta E_a^\mu = \sum_b (-1)^{F_b} \pare{ \epsilon_b'(q_*)-\epsilon_a'(p) } e^{-i q_*L} \Res_{q^1=q_*} S_{ba}^{ba} (q^1,p) \,,
\label{JL mu-term}
\end{equation}
where $q_*$ denotes the boundstate pole of $S_{ba}^{ba} (q^1,p)$.\footnote{Here we take the residue of $\mu$-term with respect to $q^1$ instead of $\tilde{q}$.}

Since we are interested in the dispersion relation of magnon excitation of $su(2|2)^2$ spin-chain \eqref{magnon dispersion:gauge}, $q^1$ is expressed in terms of $\tilde{q}$ as
\begin{equation}
q^1 = - 2i \arcsinh \(\frac{\sqrt{Q_b^2+\tilde q^2}}{4g}\) \,,
\label{mirror dispersion}
\end{equation}
where we chose the sign such that the finite-size corrections \eqref{JL F-term}, \eqref{JL mu-term} should decay at large $L$.
This dispersion relation exhibits interesting weak and strong coupling behaviors. If the virtual particle wrap $n$ times around the cylinder, its polarization effect gains the factor of $e^{-i n q^1 L}$. This factor behaves, around $\tilde q \sim 0$, as
\begin{equation}
e^{-i n q^1 L} \ \to \ \( \frac{2 g}{Q_b} \)^{2nL} \quad \( g \ll 1 \), \qquad e^{-i n q^1 L} \ \to \ \exp \( - \frac{n Q_b L}{2g} \) \quad \( g \gg 1 \).
\label{exp suppression}
\end{equation}
It shows that at strong coupling only virtual particles with $Q_b=1$ contributes to the leading exponential correction, and those with $Q_b > 1$ mix with higher wrapping effects. On the other hand, at weak coupling, virtual particles with any $Q_b$ can contribute to corrections of the order $g^{2L}$.

When the particle $a$ is a single giant magnon, one can compute its finite-size correction to the energy using the dispersion $\epsilon_a (p) \approx 4 g \abs{\sin \frac{p}{2}}$ and the $S$-matrix of the $su(2|2)^2$ spin chain \cite{Beisert05, Beisert06b, AFZ06b}. The results read \cite{JL07, HJL08}
\begin{align}
\delta E^F_{\rm GM} &= -\sqrt{\frac{g}{\pi J_1}}\, \frac{16\sin^2(\frac{p}{4})}{1-\sin(\frac{p}{2})} \, \exp \(-\frac{J_1+4g \sin(\frac{p}{2})}{2g} \) \,,
\label{eq:GM 1-loop} \\[1mm]
\delta E^\mu_{\rm GM} &= - 16g \sin^3\(\frac{p}{2}\) \, \exp \( -\frac{J_1+4g \sin(\frac{p}{2})}{2g \sin(\frac{p}{2})} \) \,.
\end{align}
The $F$-term agrees with the one-loop calculation of \cite{GSV08} and the $\mu$-term agrees with the classical energy of finite-size giant magnon \cite{AFZ06a, AFGS07}. In the previous paper \cite{HS08}, we confirmed agreement between the classical energy of finite-size dyonic giant magnon \cite{OS06} and the $\mu$-term for a magnon boundstate.

\subsection{Multi-particle states}

Let $A_\infty \equiv \{ a_1 (p_1) \cdots a_M (p_M) \}_\infty$ be a string state made up of $M$ giant magnons, with the $\ell$-th giant magnon carrying momentum $p_\ell$\,. We consider the simplest case where the polarizations of giant magnons are the same. Since giant magnons are thought of as BPS objects, the total energy of the state $A_\infty$ should be
\begin{equation}
E_{A_\infty} - J_1 = \sum_{\ell=1}^M \sqrt{ 1 + 16 g^2 \sin \frac{p_\ell}{2} } \ \approx \ \sum_{\ell=1}^M 4 g \abs{ \sin \frac{p_\ell}{2} } + \cO \( \frac{1}{g} \) \,.
\label{mgm energy1}
\end{equation}

Next, let $A \equiv \{ a_1 (p_1) \cdots a_M (p_M) \}$ be a string state made up of $M$ {\it finite-size} giant magnons. The finite-size correction $\delta E_A \equiv E_A - E_{A_\infty}$ will in general take the form
\begin{multline}
\delta E_A = \sum_{\ell =1}^M \pare{ E_0^{(\ell)} \( g, \{ p_k \} \) \, \exp \( - \frac{E_{A_\infty}}{2g \sin \frac{p_\ell}{2}} \) } + E_1 \( g, \{ p_k \} \) \, \exp \( - \frac{E_{A_\infty}}{2g} \) \\
+ \cO \( \pare{ \exp \( - \frac{E_{A_\infty}}{g \sin \frac{p_\ell}{2}} \) } , \ \exp \( - \frac{E_{A_\infty}}{g} \) \),
\end{multline}
where
\begin{align}
E_0^{(\ell)} \( g, \{ p_k \} \) &= g E_{00}^{(\ell)} \( \{ p_k \} \) + E_{01}^{(\ell)} \( \{ p_k \} \) + \cO \( \frac{1}{g} \) , 
\label{E0 series} \\
E_1 \( g, \{ p_k \} \) &= \sqrt{\frac{g}{J_1}} \, E_{10} \( \{ p_k \} \) + \cO \( \frac{1}{\sqrt g} \) .
\label{E1 series}
\end{align}
The term $E_{00}^{(\ell)} \( \{ p_k \} \)$ is the classical energy of finite-size giant magnons, and the next term $E_{01}^{(\ell)} \( \{ p_k \} \)$ is the one-loop correction to it. Origin of the term $E_{10} \( \{ p_k \} \)$ is that mode numbers of quantum fluctuation around any classical background are quantized at finite-size system, giving additional contribution to the energy.

At strong coupling, we know that $E_{00} (p) \approx \delta E_a^\mu$ and $E_{10} (p) \approx \delta E_a^F$ when $a$ is a single finite-size (dyonic) giant magnon with momentum $p = \Delta \varphi_1$\,.
Thus, we may expect the term $E_{00}^{(\ell)} \( \{ p_k \} \)$ or $E_{10} \( \{ p_k \} \)$ for multi magnon states should also match the generalized L\"uscher $\mu$- or $F$-term at strong coupling, respectively. This reasoning helps us to conjecture the generalized L\"uscher formula for multi-particle states. Our proposal is
\begin{equation}
\delta E_A (L) = \delta E_A^F (L) + \delta E_A^\mu (L)  \quad {\rm and} \quad
\delta E_A^F = \delta E_A^F {}^{\rm (main)} + \delta E_A^F {}^{\rm (back)} \,.
\label{multi formula}
\end{equation}
The $F$-terms are given by
\begin{align}
\delta E_A^F {}^{\rm (main)} &= - \sum_b (-1)^{F_b} \int_{-\infty}^{+\infty} \frac{d\tilde{q}}{2\pi} \, e^{-i q^1 L} \( \prod_{\ell=1}^M S_{b a_\ell}^{\, b a_\ell} (q^1, p_\ell)-1 \) ,
\label{multi F-term1} \\[1mm]
\delta E_A^F {}^{\rm (back)} &= + \sum_b (-1)^{F_b} \int_{-\infty}^{+\infty} \frac{d\tilde{q}}{2\pi} \pare{ \sum_{k=1}^M \alpha_k \, \frac{\epsilon_{a_k}' (p_k)}{\epsilon_b' (q^1)} } \, e^{-i q^1 L} \( \prod_{\ell=1}^M S_{b a_\ell}^{\, b a_\ell} (q^1, p_\ell)-1 \) ,
\label{multi F-term1b}
\end{align}
where $\{ \alpha_k \}$ are in general functions of $q$ and $\{ p_\ell \}$ obeying the constraint $\sum_{k=1}^M \alpha_k = 1$, and $\mu$-term is given by
\begin{equation}
\delta E_A^\mu = {\rm Re} \, \Bigg\{ \sum_{\ell=1}^M \sum_b (-1)^{F_b} \pare{\epsilon_b' (q_\ell^*) - \epsilon'_{a_\ell} (p_\ell)} e^{- i q_\ell^* L} \mathop {\rm Res} \limits_{q^1 = q_\ell^*} S_{b a_\ell}^{\, b a_\ell} (q^1 \,, p_\ell) \prod_{k \neq \ell}^M S_{b a_k}^{\, b a_k} (q_\ell^* \,, p_k) \Bigg\} ,
\label{multi mu-term1}
\end{equation}
where $q_\ell^*$ denotes the pole of $S_{b a_\ell}^{\, b a_\ell} (q^1 \,, p_\ell)$.

The object $\delta E_A^F$ consists of two terms. As will be discussed in Section \ref{sec:F-term}, the first term $\delta E_A^F {}^{\rm (main)}$ represents the energy of the virtual particle $b$, and the second term $\delta E_A^F {}^{\rm (back)}$ represents backreaction to the energy of the real particles $A$. We do not attempt to determine the backreaction part further, because it can be neglected at the leading order of approximations we will use.
Note that the backreaction part can be determined in principle for the sinh-Gordon model at finite volume discussed in Section \ref{subsec:sinh-Gordon}.

Some remarks are now in order:
\begin{itemize}
\item We have assumed the $S$-matrix to be diagonal, which is certainly the case when $a$ is a scalar and $b$ is any elementary particle of the $su(2|2)^2$ spin chain,
\begin{equation}
S (y,x) \sim S_0(y,x) \Big[ a_1 E^1_1 \otimes E^1_1 + (a_1 + a_2) E^2_2 \otimes E^1_1 + a_6 (E^3_3 \otimes E^1_1 + E^4_4 \otimes E^1_1) \Big]^2 \,.
\label{AFZ S-matrix}
\end{equation}
where $a_1, a_2, a_6$ are some functions of the momentum of particles $a$ and $b$ \cite{Beisert05, Beisert06b, AFZ06b}. This $S$-matrix \eqref{AFZ S-matrix} is called diagonal because no terms $E^i_j \otimes E^1_1 \ (i \neq j)$ are present.\footnote{Here $(E^1, E^2, E^3, E^4) = (\phi^1, \phi^2, \psi^1, \psi^2)$ signify the basis of ${\bf 2|2}$ representation, and four scalars of $\cN=4$ theory correspond to $\phi^a \bar \phi^{\dot a}$.}
When they are not diagonal, we have to modify the above formula like the conjecture of \cite{PV08}.

\item Just as in the (generalized) L\"uscher formula, our formula will be valid only when $L$ is very large. They may receive further corrections compared with TBA, such as the one coming from convolution integral, which would modify the result.

\item The expression of the $\mu$-term \eqref{multi mu-term1} can be complex-valued if we do not take its real part. To explain why we take the real part, let us recall the derivation of the generalized L\"uscher formula by Janik and \L ukowski \cite{JL07}. There they used parity symmetry of integral over $q^1$ to render $\cos (q^1 L) \mapsto 2 e^{-i q^1 L}$, which, however, may change the result when one analytically continues $q^1$ into the complex region. If we undo this procedure, we arrive at the $\mu$-term formula as shown above.

\end{itemize}

\section{The \bmt{F}-term formula for multi-particle states}\label{sec:F-term}

\subsection{One-loop finite-size correction to energy from finite-gap}\label{subsec:fg}

We start by simplifying the expression of $F$-term a little bit for the case of our concern. Since the exponential factor of \eqref{multi F-term1} is $e^{-2L \arcsinh( \sqrt{1+\tilde{q}^2}/(4g) )}$, it decays rapidly as $\tilde{q}$ increases. It allows us to evaluate the integral over $\tilde{q}$ by the saddle-point approximation for $L \gg g \gg 1$. We obtain
\begin{equation}
\delta E^F_a \approx -\int_{-\infty}^\infty \frac{d\tilde{q}}{2\pi} \; e^{-i q^1 L} \sum_b (-1)^{F_b} \prod_{\ell=1}^N S_{b a_\ell}^{\, b a_\ell}(q^1, p_\ell) \,,\qquad  q^1 \approx - i \( \frac{1}{2g} + \frac{\tilde{q}^2}{4g} \),
\label{multi F-term2}
\end{equation}
where the saddle point is at $\tilde{q}=0$, {\it i.e.} $q^1=-i/(2g)$.
The term $\sum_b (-1)^{F_b}$ disappears due to supersymmetry of the $su(2|2)^2$ spin chain, and the terms with $\alpha_\ell$ vanish because the factor $1/\epsilon_b' (q^1) \propto \epsilon_b (q^1)$ becomes zero at the saddle point. We will show that the simplified formula \eqref{multi F-term2} agrees with the exact computation of one-loop energy around finite-size multi giant magnons. Because the computation itself is straightforward extension of \cite{GSV08}, we will often skip the details.

\bigskip
As discussed in \cite{GSV08}, in order to compute the term $E_{10} \( \{ p_\ell \} \)$ of \eqref{E1 series}, we may approximate the classical background by multi giant magnons of {\it infinite} size, because this approximation just modifies the terms $E_{01}^{(\ell)}$ or higher.

Following the convention of \cite{GSV08}, we consider the simplest system of multi giant magnons, where all of them have the same polarization with the spectral parameters located outside the unit circle. The quasimomenta of such solutions are given by,
\begin{alignat}{3}
p_{\hat 1,\hat 2}(x) &= - p_{\hat 3,\hat 4}(x) & 
&= \frac{\Delta}{2 g} \frac{x}{x^2-1} \,, 
\label{multiGM qm1} \\
p_{\tilde  2}(x) &= - p_{\tilde  3}(x) & 
&= \frac{\Delta}{2 g} \frac{x}{x^2-1} + \sum_{\ell} \( \frac{1}{i}  \log\frac{x-X_\ell^+}{x-X_\ell^-} + \tilde \phi_{2, \ell} \), 
\label{multiGM qm2} \\
p_{\tilde  1}(x) &= - p_{\tilde  4}(x) & &= \frac{\Delta}{2 g} \frac{x}{x^2-1} + \sum_{\ell} \( \frac{1}{i} \log\frac{x-1/X_\ell^-}{x-1/X_\ell^+} + \tilde \phi_{1, \ell} \) ,
\label{multiGM qm3}
\end{alignat}
where $X_\ell^\pm$ are spectral parameters of the $\ell$-th (dyonic) giant magnons,
\begin{equation}
X_\ell^\pm = e^{\pm i p_\ell/2} \, \frac{ Q_\ell + \sqrt{Q_\ell^2 + 16 g^2 \sin^2 \frac{p_\ell}{2} }}{4 g \sin \frac{p_\ell}{2}} \,.
\label{spectral parameters}
\end{equation}
If we employ the orbifold regularization \cite{GSV08, RS08} to lift the constraint on the total momentum, we should choose the twists as
\begin{equation}
\tilde \phi_{1, \ell} = \tilde \phi_{2, \ell} = - p_\ell/2 ,
\label{twist choice}
\end{equation}
to meet the twisted boundary condition $\xi_1 (x=+\infty, t) = e^{i P} \, \xi_1 (x=- \infty, t),\ P \equiv \sum_\ell p_\ell$\,.

What we are going to compute as the one-loop energy is the sum over characteristic frequencies specified by the polarization $(ij)$ and the mode number $n$,
\begin{equation}
\delta \epsilon_{\text{1-loop}} = \frac12 \sum_{n \in \bb{Z}} \sum_{(ij)} (-1)^{F_{ij}} \Omega_n^{ij} \,.
\label{one-loop energy FG}
\end{equation}
Characteristic frequencies $\Omega_n^{ij}$ are usually computed by expanding the action around a given classical background \cite{FT02}. For an integrable field theory with solitons, there is an alternative way of computation, namely to add an extra small-energy soliton to a given background and compute its energy including backreaction \cite{DHN75, FK77, CDM07}. In the finite-gap language, this procedure corresponds to perturbing a given background by adding poles with appropriate residues \cite{GV07b}.

\bigskip
As shown in Appendix \ref{app:FG}, on the multi giant magnon background \eqref{multiGM qm1}-\eqref{multiGM qm3}, the characteristic frequencies $\Omega_n^{ij}$ become
\begin{equation}
\Omega_n^{ij} = \Omega (x_n^{ij}), \qquad
\Omega (x) = \frac{2}{x^2-1} \[ 1 - \sum_{\ell=1}^M \alpha_\ell \(\frac{X_\ell^- + X_\ell^+}{X_\ell^- X_\ell^+ +1}\) x \], \qquad
\sum_{\ell=1}^M \alpha_\ell = 1.
\label{M-mag Omega}
\end{equation}
To understand the meaning of \eqref{M-mag Omega}, let us recall that if we evaluate the energy of a giant magnon \eqref{magnon dispersion:string} in the plane-wave limit $\ln \( X_{pw}^+/X_{pw}^- \) \equiv i Q_{pw} P_{pw}/(2g) \ll 1$, we obtain the dispersion relation $E_{pw} = \sqrt{\mathstrut{} 1 + P_{pw}^2}$\,. If we parametrize the energy and the momentum of plane-waves by
\begin{equation}
E_{pw} = \frac{x^2 + 1}{x^2 - 1} = 1 + \epsilon_{\rm fluc} \,, \quad P_{pw} = \frac{2 x}{x^2 - 1} = p_{\rm fluc} \,,
\label{planewave dispersion}
\end{equation}
the function $\Omega (x)$ can be rewritten as
\begin{equation}
\Omega (x) = \epsilon_{\rm fluc} - \sum_{\ell=1}^M \alpha_\ell \( \frac{d \epsilon_{\ell, \infty} }{d p_\ell} \) p_{\rm fluc} 
\equiv \epsilon_{\rm fluc} + \sum_{\ell=1}^M \delta p_\ell \( \frac{d \epsilon_{\ell, \infty} }{d p_\ell} \).
\label{M-mag Omega2}
\end{equation}
Thus, the function $\Omega (x)$ roughly stands for the energy of plane-wave excitations and its backreaction onto multi giant magnons weighted by $\alpha_\ell$\,, as discussed in \cite{GSV08}. From \eqref{M-mag Omega2} we can also find the conservation of momentum
\begin{equation}
\sum_{\ell=1}^M \delta p_\ell + p_{\rm fluc} = \( - \sum_{\ell=1}^M \alpha_\ell + 1 \) p_{\rm fluc} = 0.
\label{FG conservation}
\end{equation}

\bigskip
Once the function $\Omega (x)$ is known, one can carry out the summation over mode numbers and polarizations as in \cite{GSV08}. At an intermediate step, one can show $\sum_{ij} (-1)^F \( p_i - p_j \) = 0$ for the multi giant magnon background \eqref{multiGM qm1}-\eqref{multiGM qm3}, which shows the one-loop energy of such background indeed vanishes at infinite size as in \eqref{mgm energy1}. The first nontrivial term is
\begin{equation}
\delta \epsilon_{\text{1-loop}} \approx \oint_{\bb{U}_+} \frac{dx}{2 \pi i}\, \sum_{(ij)}(-1)^{F_{ij}} \, e^{- i (p_i-p_j)} \partial_x \Omega(x) ,
\label{1-loop x-integral}
\end{equation}
where $\bb{U}_+$ is the upper half part of the unit circle. The sum over polarizations becomes
\begin{equation}
\sum_{(ij)} (-1)^{F_{ij}}e^{-i(p_i-p_j)} = \(A^+ + A^- - 2 \)^2 \exp\(-\frac{i \Delta}{g} \frac{x}{x^2-1}\) \,,
\label{flavor sum GM2}
\end{equation}
where
\begin{equation}
A^+ = \prod_{\ell} \frac{x-X_\ell^-}{x-X_\ell^+} \, e^{-i \tilde \phi_{2, \ell}} \,, \quad
A^- = \prod_{\ell} \frac{x-1/X_\ell^+}{x-1/X_\ell^-} \, e^{-i \tilde \phi_{1, \ell}} \,, \quad
\Delta = J_1 + \sum_\ell \sqrt{Q_\ell^2+16g^2 \sin^2 \frac{p_\ell}{2}} .
\label{flavor sum GM3}
\end{equation}
We find approximate equality $A^+ \approx A^-$ from the choice of twist \eqref{twist choice} and $X_\ell^+ X_\ell^- \approx 1$ coming from \eqref{spectral parameters} when $Q_\ell$ is small. 
However it follows $X_\ell^+ X_\ell^- > 1$ for $Q_\ell \gg 1$, so $A^+$ and $A^-$ are no longer approximately equal.

The other factor in the integrand, $\partial_x \Omega (x)$, is proportional to the `kinematical factor' of the $F$-term formula,
\begin{equation}
\partial_x \Omega(x) = - \frac{4 x}{(1 - x^2)^2} \( 1 - \sum_\ell \alpha_\ell \, \epsilon_{a_\ell}' (p_\ell) \, \frac{1+x^2}{2x} \) \ \propto \ \( 1 - \sum_\ell \alpha_\ell \frac{\epsilon_{a_\ell} '(p_\ell)}{\epsilon_b '(q^1)} \),
\label{dOmega dx}
\end{equation}
where $\epsilon_b' (q^1) \equiv \epsilon_{\rm fluc}' (p_{\rm fluc})$.

\bigskip
Finally we evaluate the integral \eqref{1-loop x-integral} by saddle-point approximation. As one can see from \eqref{flavor sum GM2}, the saddle point lies at $x = i$, which implies $E_{\rm fluc} (p_{\rm fluc}) \propto 1/\epsilon_b' (q^1) = 0$ and the terms proportional to $\alpha_\ell$ in \eqref{dOmega dx} drop off. Moreover, the sum over flavor \eqref{flavor sum GM2} can be identified with the $S$-matrix part of the $F$-term formula:
\begin{align}
\sum_{(ij)} (-1)^{F_{ij}}e^{-i(p_i-p_j)} &\approx 4 \prod_{\ell=1}^M S_0 (x_q , x_{p_\ell}) \( \prod_{\ell=1}^M a_1 (x_q , x_{p_\ell} ) - \prod_{\ell=1}^M a_6 (x_q , x_{p_\ell} ) \)^2  \notag \\
&= \sum_b (-1)^{F_b} \prod_{\ell=1}^M S_{b a_\ell}^{\, b a_\ell} (q^1, p_\ell).
\label{S-matrix sum}
\end{align}
To show these equalities in the case of multi giant magnons, we have to use the following kinematics
\begin{equation}
x_{p_\ell}^\pm = e^{\pm i p_\ell/2} \,, \qquad x_q^\pm = i, \quad \( q^1 = - \frac{i}{2g} \),
\label{saddle point xpxq}
\end{equation}
as well as the expressions of $su(2|2)^2\ S$-matrix at strong coupling \cite{AFZ06b, Beisert05, Beisert06b},
\begin{alignat}{3}
S_0 (x_q , x_p ) &\equiv \frac{x_q^{-} - x_p^{+}}{x_q^{+} - x_p^{-}} \, \frac{1-\frac{1}{x_p^{-} x_q^{+}}}{1-\frac{1}{x_p^{+} x_q^{-}}} \, \sigma^2(x_q,x_p) &\; &\approx e^{-2 \sin \frac{p}{2}} \,,
\label{s0qp} \\
a_1 (x_q \,, x_p) &\equiv \frac{x_p^{-} - x_q^{+}}{x_p^{+} - x_q^{-}} \, \frac{\eta (x_p) \eta (x_q)}{\tilde \eta (x_p) \tilde \eta (x_q)} &\; &\approx \frac{e^{-ip/2} - i}{e^{ip/2} - i} \, e^{ip/2} \,, 
\label{a1qp} \\
a_2 (x_q \,, x_p) &\equiv \frac{(x_q^{-} - x_q^{+})(x_p^- - x_p^+)(x_p^- + x_q^+)}{(x_q^- - x_p^+)(x_p^- x_q^- - x_p^+ x_q^+)} \, \frac{\eta (x_p) \eta (x_q)}{\tilde \eta (x_p) \tilde \eta (x_q)} &\; &\approx {\cal O} (g^{-1}) \,,
\label{a2qp} \\
a_6 (x_q \,, x_p) &\equiv \frac{x_q^+ - x_p^+}{x_q^- - x_p^+} \, \frac{\eta (x_p)}{\tilde \eta (x_p)} &\; &\approx 1 \,,
\label{a6qp}
\end{alignat}
with the choice of the so-called string frame $\eta (x_q)/\tilde \eta (x_q) = \sqrt{ x_p^+/x_p^- },\ \eta (x_p)/\tilde \eta (x_p) = \sqrt{ x_q^-/x_q^+ }$.

For multi dyonic giant magnons, we should use the elementary-boundstate $S$-matrix given in \cite{HS08}. By using the same saddle point approximation, we obtain in the string frame,
\begin{alignat}{3}
\sum_{b} &(-1)^{F_b} S_{bA}^{bA} (x_q, \{ X_\ell \}) \Bigr|_{x_q^\pm \approx i} \notag \\
&= \prod_{\ell=1}^M S_{\rm BDS} (x_q, X_\ell) \, \sigma^2 (x_q, X_\ell) \( \frac{\eta (X_\ell)}{\tilde \eta (X_\ell)} \)^2 \( \frac{\eta (x_q)}{\tilde \eta (x_q)} \)^{2 Q_\ell} \times \notag \\
&\hspace{70mm} \[ 1 + \prod_{\ell=1}^M s_2 (x_q, X_\ell) - 2 \prod_{\ell=1}^M s_3 (x_q, X_\ell) \]^2 \Bigr|_{x_q^\pm \approx i}  \notag \\[1mm]
&\approx \[ \prod_{\ell=1}^M \frac{i-X_\ell^-}{i-X_\ell^+} \, e^{i p_\ell/2} + \prod_{\ell=1}^M \frac{i-1/X_\ell^+}{i-1/X_\ell^-} \, e^{i p_\ell/2} - 2 \]^2 \prod_{\ell=1}^M \exp \( - \frac{\epsilon_{a_\ell} (p_\ell)}{2g} \),
\end{alignat}
which agrees with \eqref{flavor sum GM2}.

In summary, collecting the results \eqref{1-loop x-integral}, \eqref{flavor sum GM2}, \eqref{flavor sum GM3}, \eqref{dOmega dx} and \eqref{S-matrix sum}, one can find that one-loop finite-size correction to the energy of multi (dyonic) giant magnons agrees with the generalized $F$-term formula for multi-particle states advertised in \eqref{multi F-term2}.

\subsection{One-loop finite-size correction to the energy of multi giant magnons}

Here we explicitly evaluate the one-loop finite-size correction to the state with $M$ giant magnons, $A = \{ a (p_1) \ldots a_M (p_M) \}$.

From the argument at the beginning of Section \ref{subsec:fg}, the $F$-term for the $M$-magnon state reduces to
\begin{equation}
\delta E_A^F=-\int_{-\infty}^\infty \frac{d\tilde{q}}{2\pi} \, e^{-L(\frac{1}{2g}+\frac{\tilde{q}^2}{4g})} \cdot 4 \prod_{\ell=1}^M S_0 (x_q , x_{p_\ell}) \( \prod_{\ell=1}^M a_1 (x_q , x_{p_\ell} ) - \prod_{\ell=1}^M a_6 (x_q , x_{p_\ell} ) \)^2,
\end{equation}
at strong coupling. Performing the integral and using \eqref{s0qp}-\eqref{a6qp}, we obtain
\begin{equation}
\delta E_A^F = -4\sqrt{\frac{g}{\pi L}} \( \prod_{\ell=1}^M \frac{\cos(\frac{p_\ell}{4})+\sin(\frac{p_\ell}{4})}{\cos(\frac{p_\ell}{4})-\sin(\frac{p_\ell}{4})}-1 \)^2
e^{-\frac{1}{2g}\Bigl(L+\sum\limits_{\ell=1}^M 4g\sin(\frac{p_\ell}{2})\Bigr)}\,.
\end{equation}
Interestingly the exponential part is generalization of \eqref{eq:GM 1-loop}.

If we choose the so-called spin-chain frame $\eta (x_q)/\tilde \eta (x_q) = \eta (x_p)/\tilde \eta (x_p) = 1$, the $F$-term becomes
\begin{equation}
\delta E_{\text{spin-chain}}^F=-4\sqrt{\frac{g}{\pi L}} \( e^{-iP/2} \prod_{\ell=1}^M \frac{\cos(\frac{p_\ell}{4})+\sin(\frac{p_\ell}{4})}{\cos(\frac{p_\ell}{4})-\sin(\frac{p_\ell}{4})}-1 \)^2
e^{-\frac{1}{2g}\Bigl(L+\sum\limits_{\ell=1}^M 4g\sin(\frac{p_\ell}{2})\Bigr)}\,,
\end{equation}
where $P=\sum_{\ell} p_\ell$ is the total momentum.

\subsection{Comparison with exact results in sinh-Gordon model}\label{subsec:sinh-Gordon}

In this subsection, we compare our proposed $F$-term with the results of sinh-Gordon model in finite volume, which is solved exactly in \cite{Teschner07}.
The finite volume spectrum for $M$-particle state in sinh-Gordon model are determined by the following non-linear integral equations \cite{Teschner07},
\begin{align}
E(L) &= \sum_{j=1}^M m \cosh \theta_j
-\int_{-\infty}^\infty \frac{d\theta}{2\pi} \, m \cosh \theta \, K(\theta),
\label{eq:E} \\
\log Y(\theta) &= -mL\cosh\theta -\sum_{j=1}^M \log S(\theta-\theta_j-i\frac{\pi}{2})+\sigma*K(\theta), 
\label{eq:logY}
\end{align}
where $S(\theta)$ is the $S$-matrix which is given by
\begin{equation}
S(\theta) = \frac{\sinh \theta-i\sin(\theta_0)}{\sinh\theta+i \sin(\theta_0)}
\;\;\;{\rm with} \;\;\; \theta_0 = \frac{\pi b^2}{1+b^2}\,,
\end{equation}
and $\sigma(\theta),\,K(\theta)$, convolution integral $f*g(\theta)$ are defined as follows:
\begin{equation}
\sigma(\theta) = -i\frac{d}{d \theta}\log S(\theta),\quad
K(\theta) = \log(1+Y(\theta)) ,\quad
f*g(\theta) = \int_{-\infty}^\infty \frac{d\theta'}{2\pi} f(\theta-\theta')g(\theta')\,.
\end{equation}
The parameters $\theta_j (j=1,\dots,M)$ are determined by the equation $\log Y(\theta_j+i\pi/2)=(2n_j+1)i\pi$ where $n_j$ is an integer.

For a large $L$ case, we can neglect the convolution term because $K(\theta)={\cal O}(e^{-mL\cosh\theta})$. In this case, since $\log Y(\theta_j+i\pi/2)=(2n_j+1)i\pi$, the equation \eqref{eq:logY} reduces to the Bethe Ansatz equations
\begin{equation}
e^{-imL \sinh\theta_j} = \prod_{k \neq j}^M S(\theta_j-\theta_k).
\label{eq:BAE}
\end{equation}
Substituting \eqref{eq:logY} to the second term of \eqref{eq:E}, we find that the last term of \eqref{eq:E} has a suggestive expression:\footnote{The second part of $\prod S-1$ comes from the groundstate finite-size correction.}
\begin{equation}
\delta E_{\rm shG}^{\rm (main)} (L)=-\int_{-\infty}^\infty \frac{d\theta}{2\pi} \, m \cosh\theta \, e^{-mL\cosh\theta} \prod_{j=1}^M S(\theta_j-\theta+i\frac{\pi}{2}).
\end{equation}
One can show that this result can be obtained from the main part of the generalized $F$-term formula \eqref{multi F-term1}, by substituting
\begin{gather}
\tilde{q} = iq^0 = im \cosh \(\theta - \frac{i \pi}{2}\) = m\sinh \theta, \quad
q^1 = m \sinh \(\theta - \frac{i \pi}{2}\) = - im \cosh \theta,
\notag \\
S_{b a_j}^{\, b a_j} (x_q, x_{p_j}) = S (\theta_j - \theta + \frac{i\pi}{2}),
\end{gather}
and neglecting the backreaction term $\alpha_\ell$. Both results agree at leading order of saddle-point approximation.

The backreaction part of the $F$-term seems to correspond to the convolution part of \eqref{eq:logY}. After including the convolution, the Bethe Ansatz equation \eqref{eq:BAE} is modified as
\begin{align}
\log Y\(\theta_j+i\frac{\pi}{2}\) &= (2n_j+1)i\pi \notag \\
&= - imL\sinh\theta_j - i\pi - \sum_{k \neq j}^M \log S(\theta_j-\theta_k) + \sigma*K \(\theta_j+\frac{\pi}{2}\),
\label{eq:BAE_mod}
\end{align}
where for large $L$ the last term becomes,
\begin{equation}
\sigma*K\(\theta_j+i\frac{\pi}{2}\)
\approx \int_{-\infty}^\infty \frac{d\theta}{2\pi} \, \sigma\(\theta_j+i\frac{\pi}{2}-\theta\) e^{-mL\cosh \theta} \prod_{k=1}^M S(\theta-\theta_k+i\frac{\pi}{2}) \,.
\label{large-L convolution}
\end{equation}
The parameters $\{ \theta_j \}$ receive corrections due to the convolution. We write these corrections as $\theta_j=\tilde{\theta}_j+\delta \theta_j$
where $\{ \tilde{\theta}_j \}$ are the solutions of the equations \eqref{eq:BAE},
\begin{equation}
e^{-imL \sinh\tilde{\theta}_j} = \prod_{k \neq j}^M S(\tilde{\theta}_j-\tilde{\theta}_k)\,.
\label{eq:BAE2}
\end{equation}
Substituting $\theta_j=\tilde{\theta}_j+\delta \theta_j$ into \eqref{eq:BAE_mod} and using \eqref{eq:BAE2}, one can determine the form of $\delta \theta_j$ as
\begin{equation}
\delta \theta_j = \int_{-\infty}^\infty \frac{d\theta}{2\pi} \, A_j (\{\tilde{\theta_\ell}\},\theta) \,
e^{-mL\cosh \theta} \prod_{k=1}^M S(\theta - \tilde \theta_k+i\frac{\pi}{2}),
\end{equation}
at the leading order of $e^{-mL\cosh \theta}$. 
The coefficients $A_j(\{\tilde{\theta_\ell}\},\theta)$ follow from an infinitesimal variation of the equation \eqref{eq:BAE_mod}, though their actual expression will be complicated. Note that $n_j$'s are quantized at integers and cannot be varied. At large $L$, one can deduce a sum rule among $\delta \theta_j$ from \eqref{eq:BAE_mod} and \eqref{large-L convolution}, which reads
\begin{align}
\delta P_{\rm total} &\equiv m \sum_{j=1}^M \cosh \tilde \theta_j \, \delta \theta_j  \notag \\
&= \frac{1}{iL} \sum_{j=1}^M
\int_{-\infty}^\infty \frac{d\theta}{2\pi} \, \sigma\(\tilde \theta_j + i\frac{\pi}{2} - \theta\) e^{-mL\cosh \theta} \prod_{k=1}^M S(\theta - \tilde \theta_k+i\frac{\pi}{2}) \,.
\label{momentum violation}
\end{align}
This relation is analogous to \eqref{FG conservation}, but the total momentum defined as above is not kept fixed once the convolution is taken into account.

Substituting these results into \eqref{eq:E} once again, we obtain
\begin{equation*}
E(L) =\sum_{j=1}^M m\cosh \tilde{\theta}_j+\sum_{j=1}^M m\sinh \tilde{\theta}_j \delta \theta_j
-\int_{-\infty}^\infty \frac{d\theta}{2\pi} m \cosh \theta K(\theta) \equiv \sum_{j=1}^M m\cosh \tilde{\theta}_j+\delta E_{\rm shG} (L)\,,
\end{equation*}
where
\begin{align*}
\delta E_{\rm shG} (L) = -\int_{-\infty}^\infty \frac{d\theta}{2\pi}
\(m\cosh\theta-\sum_{j=1}^M  m \sinh \tilde{\theta}_j
A_j(\{\tilde{\theta_\ell}\},\theta)\) e^{-mL\cosh \theta} \prod_{k=1}^M S(\tilde \theta_k-\theta+i\frac{\pi}{2})\,.
\end{align*}
This result is consistent with our proposal for the generalized $F$-term for multi-particle states \eqref{multi F-term1} and \eqref{multi F-term1b}.

\section{The \bmt{\mu}-term formula for multi-particle states}\label{sec:mu-term}

As explained in Section \ref{sec:luscher}, the $\mu$-term for one-particle states arises from an on-shell splitting process $a_\ell (p_\ell) \to b (q_b) + c (q_c)$ which corresponds to the boundstate pole of the $S$-matrix between $a_\ell$ and $b$.
Similarly, we expect the generalized $\mu$-term for multi-particle states admits a similar interpretation. If the $F$-term integral picks up a pole of the $S$-matrix $S_{b a_\ell}^{\, b a_\ell}$ and if $\alpha_k = \delta_{k \ell}$ holds, that is if the backreaction localizes around the $\ell$-th soliton, this contribution is written as
\begin{equation}
\delta E_A^\mu = \sum_{\ell=1}^M \sum_b (-1)^{F_b} \pare{\epsilon_b' (q_\ell^*) - \epsilon'_{a_\ell} (p_\ell)} e^{- i q_\ell^* L} \mathop {\rm Res} \limits_{q^1 = q_\ell^*} S_{b a_\ell}^{\, b a_\ell} (q^1 \,, p_\ell) \prod_{k \neq \ell}^M S_{b a_k}^{\, b a_k} (q_\ell^* \,, p_k).
\label{multi mu-term2}
\end{equation}
If we take the real part of this equation, we obtain our $\mu$-term formula \eqref{multi mu-term1}.

We are going to consider two examples of multi-magnon states in order to give support for our conjecture on the generalized $\mu$-term \eqref{multi mu-term1}. The first example is the state composed of several giant magnons, and the second example is the state composed of several dyonic giant magnons where each dyonic giant magnon carries a large angular momentum in the second direction. Finite-size corrections to the energy of these states have been computed in \cite{MO08}. Indeed, this result is correctly reproduced from our $\mu$-term formula as we show in Section \ref{sec:classical mgm}.

We may also consider solving the asymptotic Bethe Ansatz in the $su(2)$ sector at a finite length. Recall that the finite-size correction to the energy of a single dyonic giant magnon was computed by using the generalized L\"uscher formula in \cite{HS08} and by solving the Bethe Ansatz in the $su(2)$ sector \cite{MO08}, and the two results turned out to coincide.
It turns out that this coincidence remains approximately valid for certain multi-particle states.

In general, however, these two methods do not give the same answer. For example, the finite-size correction to the energy of a single giant magnon is computed by the generalized L\"uscher formula \cite{JL07} and that of a two-magnon boundstate is computed by the asymptotic Bethe Ansatz \cite{AF07}.
It turns out that only the former computation is consistent with the result of string theory \cite{AFZ06a, AFGS07}. Note that this mismatch by itself is not surprising because the asymptotic Bethe Ansatz may receive corrections at finite size, as argued in \cite{AF07}.

It should be kept in mind that a similar idea has already been pointed out by Pozsgay \cite{Pozsgay08}. He observed that the computation of the finite-size correction to energy based on Bethe Ansatz equations resembles the L\"uscher's $\mu$-term formula for relativistic integrable theories, both of which exhibit exponential suppression in size. So our strategy may be regarded as generalization of Pozsgay's analysis to non-relativistic theories.

\subsection{Classical finite-size correction to the energy of multi giant magnons}\label{sec:classical mgm}

Finite-size corrections to the classical energy of multi giant magnons have been computed by Minahan and Ohlsson-Sax using the finite-gap technique in \cite{MO08}. In the finite-gap language, a giant magnon or a dyonic giant magnon is expressed as a condensate, namely a segment running from $x=X_j^+$ to $x=X_j^-$ with constant density \cite{MTT06}. The state of $M$ (dyonic) giant magnons is expressed by $M$ condensates. When the total angular momentum becomes finite, small branch cuts evolve from the endpoints of the condensates and the positions of the endpoints are also slightly shifted. Taking these effects into account, they computed finite-size correction to the energy of multi giant magnons at the leading order of $e^{-J_1}$ which is given by
\begin{align}
\Delta E &= g^2 \sum_{\ell=1}^M \frac{\sin^4 \frac{p_\ell}{2}}{\epsilon_{Q_\ell} (p_\ell) } \, {\rm Re} \[ \( \frac{\delta_\ell \ssp e^{i \phi_\ell}}{X_\ell^+} \)^2 e^{-i p_\ell} \], 
\label{fs-mdgm energy} \\[2mm]
\frac{\delta_\ell \ssp e^{i \phi_\ell}}{X_\ell^+} &= 8 \exp \[ -\frac{i E}{4} \( \frac{1}{X_\ell^+ + 1} + \frac{1}{X_\ell^+ - 1} \) + i \pi n_\ell \] \, \prod_{k \neq \ell}^M \frac{X_\ell^+ - X_k^-}{X_\ell^+ - X_k^+} \,,
\label{delta phi l}
\end{align}
where $E = J_1 + \sum_{k=1}^M \epsilon_{Q_k} (p_k)$ is the total energy of multi giant magnons at infinite size and $n_\ell$ are integers.

\bigskip
When there is only one (dyonic) giant magnon, this result \eqref{fs-mdgm energy} agrees with that of the generalized L\"uscher $\mu$-term formula \cite{JL07, HS08}. We will see below that the agreements continue to hold for the states made up of many (dyonic) giant magnons.

Let us first consider a string state made up of many single-spin giant magnons, $A = \{ a(p_1) \ldots a (p_M) \}$. The spectral parameters of each giant magnon are written as $X_j^\pm \approx e^{\pm i p_j/2}$. By substituting the $S$-matrix \eqref{AFZ S-matrix} with the coefficients \eqref{s0qp}-\eqref{a6qp} into the $\mu$-term \eqref{multi mu-term1} and taking the strong coupling limit $g \gg 1$, we obtain
\begin{multline}
\delta E^\mu_A = {\rm Re} \Biggl\{ \sum_{\ell=1}^M \pare{ \epsilon'(q_\ell^*)-\epsilon'(p_\ell)} e^{-iq_\ell^* L} \cdot 4 \Res_{q^1=q_\ell^*} \( a_1(x_q,x_{p_\ell}) \)^2 S_0(x_q,x_{p_\ell})  \\
\times \prod_{k \neq \ell}^M (a_1( x_{q_\ell^*}, x_{p_k} ))^2 S_0( x_{q_\ell^*}, x_{p_k}  ) \Biggr\} .
\end{multline}
where the momentum $q_\ell^*$ is determined from the physical pole of $S^{ba_\ell}_{ba_\ell}(q^1,p_\ell)$, as
\begin{equation}
q_\ell^*=-\frac{i}{2g \sin(\frac{p_\ell}{2})} \,.
\end{equation}
Now we can borrow the result for one-particle state in \cite{JL07},
\begin{equation}
\pare{ \epsilon'(q_\ell^*) - \epsilon'(p_\ell)} e^{-iq_\ell^* L} \cdot 4 \Res_{q=q_\ell^*} \( a_1(x_{p_\ell},x_q) \)^2 S_0(x_{p_\ell},x_q)
= - 16g \sin^3 \left( \frac{p_\ell}{2} \right) e^{- 2 - \frac{L}{2g \sin(\frac{p_\ell}{2})}}.
\end{equation}
The final result of the $\mu$-term is, assuming no two momenta are equal $(p_j \neq p_k$ for $j \neq k)$,
\begin{equation}
\delta E^\mu_A = \sum_{\ell=1}^M \left( -16g \sin^3 \left( \frac{p_\ell}{2} \right) \prod_{k \neq \ell}^M \frac{\sin^2(\frac{p_\ell+p_k}{4})}{\sin^2(\frac{p_\ell-p_k}{4})} \exp \left[ -\frac{L+\sum_{k=1}^M 4g \sin(\frac{p_k}{2})}{2g \sin(\frac{p_\ell}{2})} \right] \right) .
\end{equation}
which is exactly identical to the result of \cite{MO08} by identifying $L$ with $J_1$.

Next, we consider the state made up of $M$ dyonic giant magnons with the $j$-th dyonic giant magnon carrying the second angular momentum $Q_j \gg 1$\,; $A = \{ A_1 (p_1) \ldots A_M (p_M) \}$. We write the spectral parameters of each dyonic giant magnon as $X_j^\pm \approx e^{(\pm i p_j - \theta_j)/2}$. In this case, the generalized $\mu$-term formula at strong coupling is slightly modified to
\begin{multline}
\delta E^\mu_A = {\rm Re} \Biggl\{ \sum_{\ell=1}^M \pare{ \epsilon'(q_\ell^*)-\epsilon_{Q_\ell}'(p_\ell)} e^{-iq_\ell^* L} \cdot 2 \Res_{q^1=q_\ell^*} \( A_1(x_q,X_{p_\ell}) \)^2 S_0(x_q,X_{p_\ell})  \\
\times \prod_{k \neq \ell}^M (A_1( x_{q_\ell^*}, X_{p_k} ))^2 S_0( x_{q_\ell^*}, X_{p_k}  ) \Biggr\} .
\label{generalized mu-term bound}
\end{multline}
where $A_1 (Y, X_\ell) = a_1(Y,x_1) a_1(Y,x_2) \cdots a_1 (Y,x_{Q_\ell})$ is the elementary-boundstate $S$-matrix in the $su(2)$ subsector, and the momentum $q_\ell^*$ is determined by
\begin{equation}
q_\ell^*=-\frac{i}{2g \sin(\frac{p_\ell - i \theta_\ell}{2})} \,, \qquad {\rm or} \qquad x^+ (q_\ell^*) \approx x^- (q_\ell^*) = X^+ (p_\ell) \equiv X_\ell^+ \,.
\label{generalized mu-pole bound}
\end{equation}
We multiplied a factor of 2 in \eqref{generalized mu-term bound} to take into account the contributions from two poles $x_q^\pm = X_\ell^+$ \cite{HS08}.

The first line of \eqref{generalized mu-term bound} is much the same as the $M=1$ case, so we reuse the result of \cite{HS08}. In the strong coupling limit, the dressing phase appearing in the second line becomes
\begin{equation}
\sigma^2 (x_{q_\ell^*}, X_{p_k}) \approx \( \frac{1 - \frac{1}{X_\ell^+ X_k^+}}{1 - \frac{1}{X_\ell^+ X_k^-}} \)^2 \exp \( -\frac{\epsilon_{Q_k} (p_k) - Q_k}{2 g \sin \(\frac{p_\ell - i \theta_\ell}{2} \)} \),
\end{equation}
and it follows that
\begin{equation}
\prod_{k \neq \ell}^M A_1( x_{q_\ell^*}, X_{p_k} )^2 S_0 (x_{q_\ell^*}, x_{p_k}) \approx \prod_{k \neq \ell}^M \( \frac{X_\ell^+ - X_k^-}{X_\ell^+ - X_k^+} \)^2 \exp \( - \frac{\epsilon_{Q_k} (p_k)}{2 g \sin \(\frac{p_\ell - i \theta_\ell}{2} \)} + i p_k \),
\end{equation}
where we chose the string frame. From the momentum conservation, we find $\sum_{k \neq \ell} \ssp (i p_k) = 2 \pi i n_\ell - i p_\ell$\,. Thus, assuming no two momenta are equal $(X_j^\pm \neq X_k^\pm$ for $j \neq k)$, the $\mu$-term for multi dyonic giant magnons becomes
\begin{equation}
\delta E_A^\mu = {\rm Re} \pare{ \sum_{\ell=1}^M \Biggl( - 4 \ssp g \, \frac{\sin^4 \(\frac{p_\ell}{2}\)}{\epsilon_{Q_\ell} (p_\ell)} \, e^{- i p_\ell} \prod_{k \neq \ell}^M \( \frac{X_\ell^+ - X_k^-}{X_\ell^+ - X_k^+} \)^2 \exp \[ - \frac{L + \sum_{k=1}^M \epsilon_{Q_k} (p_k)}{2 g \sin \( \frac{p_\ell - i \theta_\ell}{2} \)} \] } \Biggr).
\end{equation}
This agrees with the result of finite-gap analysis \eqref{fs-mdgm energy} by identifying $L \leftrightarrow J_1$\,.

\subsection{The \bmt{\mu}-term from the asymptotic Bethe Ansatz}

Results of the last subsection convince ourselves that our $\mu$-term formula for multi-particle states is correct. Below we will reconsider the case where all particles are boundstates of a large number of elementary magnons. In this situation, the results of generalized $\mu$-term formula become approximately equal to those of the asymptotic Bethe Ansatz at finite but large $L$, where $L$ is the length (or size) of spin chain.

In the rest of this section, we will use the notation summarized in Appendix \ref{sec:notation}.

\subsubsection{One-particle states}

For the moment we focus on a single magnon boundstate. Let $A = \{a_1 (p_1) \cdots a_Q (p_Q)\}$ be a $Q$-magnon boundstate with $Q>1$. For simplicity we assume the state $A$ belongs to the $su(2)$ sector. The $S$-matrix is of the form \cite{BDS04, AFS04}
\begin{align}
S_L (u_i \,, u_j) &= \frac{u_i - u_j + i}{u_i - u_j - i} \, e^{i \varphi_L (u_i \,, u_j)} \,,
\label{multi-L S-matrix assumed} \\
u_j = u (p_j) &= \frac12 \cot \(\frac{p_j}{2}\) \sqrt{Q_j^2+16g^2\sin^2 \(\frac{p_j}{2}\)},
\label{multi-L rapidity}
\end{align}
with $u_i - u_j = i$ being a physical boundstate pole. We also assume that the phase $\varphi_L (u_i \,, u_j)$ is independent of the spin chain size $L$ approximately up to small exponential corrections. The periodic boundary condition takes the usual form
\begin{equation}
1 = e^{- i p_j L} \prod_{i \neq j} S_L (u_j \,, u_i), \qquad P \equiv \sum_{j=1}^Q p_j = \frac{2 \pi I}{L} \quad \( I \in \bb{Z} \).
\label{mu-term pbc1}
\end{equation}
It then immediately follows that when $L=\infty$ Bethe roots of a $Q$-particle boundstate spread as
\begin{equation}
u_{j, \infty} = U_\infty + i \( j - \frac{Q+1}{2} \), \qquad (j=1,2, \ldots , Q),
\label{Q-bound rapidity}
\end{equation}
where $U_\infty$ is real so that the energy of the boundstate be real.

From the Bethe Ansatz equations \eqref{mu-term pbc1}, we expect that when $L$ is finite the rapidities of constituent magnons be displaced by $u_j = u_{j, \infty} + \Delta u_j$\,.
As discussed in \cite{MO08}, the Bethe Ansatz equations \eqref{mu-term pbc1} tell us that the difference of the displacements $\Delta u_j$ among the inner momenta are negligibly small:
\begin{gather}
\begin{array}{*9{l}}
1 &\gg &\Delta u_1 - \Delta u_2 &\gg &\Delta u_2 - \Delta u_3 &\gg &\cdots , \\
1 &\gg &\Delta u_Q - \Delta u_{Q-1} &\gg &\Delta u_{Q-2} - \Delta u_{Q-3} &\gg &\cdots .
\end{array}
\label{outermost dominance}
\end{gather}
Thus at leading order, we can write the displacement of rapidities as
\begin{equation}
\Delta u_1 = \Delta U + \Delta \tilde u_1 \,, \quad
\Delta u_k = \Delta U \ \ \(k = 2, \ldots , Q-1\), \quad
\Delta u_Q = \Delta U + \Delta \tilde u_Q \,.
\label{outermost dominance2}
\end{equation}
For later purpose, let us rewrite these displacements in terms of momentum,
\begin{equation}
\Delta p_1 = \Delta p_{1, U} + \Delta \tilde p_1 \,, \quad
\Delta p_k = \Delta p_{k, U} \ \ \(k = 2, \ldots , Q-1 \), \quad 
\Delta p_Q = \Delta p_{Q, U} + \Delta \tilde p_Q  \,, 
\end{equation}
where $\Delta p_{j, U}$ accounts for the overall shift of rapidity $\Delta U$.

Since the mode number $I$ in \eqref{mu-term pbc1} is an integer, it must be invariant as we vary $L$. It then results in the conservation of the total momentum $\Delta P/P = \Delta L/L \approx 0$, if $L$ is sufficiently large.
From this momentum conservation it follows that
\begin{equation}
0 \approx \Delta P = \Delta p_1 + \sum_{k=2}^{Q-1} \Delta p_k + \Delta p_Q \ =\ \Delta \tilde p_1 + \sum_{j=1}^Q \Delta p_{j, U} + \Delta \tilde p_Q \,.
\label{multi-L Delta P}
\end{equation}
Let us denote the sum $\sum_j \Delta p_{j, U}$ by $\Delta P_U$\,. Since this is the displacement of the boundstate momentum $P$ caused by $\Delta U$, we should have the relation $\Delta U = (\Delta P_U) \ssp U'(P)$ where the function $U(P)$ is given in \eqref{def:rapidity U}.

Substituting the parametrization \eqref{outermost dominance2} into the Bethe Ansatz equations \eqref{mu-term pbc1}, we can express the displacements $\Delta \tilde u_1$ and $\Delta \tilde u_Q$ as functions of rapidities at $L=\infty$. Since we assumed the difference between $S_L$ and $S_\infty$ is negligible, we obtain
\begin{equation}
\Delta \tilde u_1 = e^{i p_{1, \infty} L} \mathop {\rm Res} \limits_{u_1 = u_{1, \infty}} \prod_{i=2}^Q S_\infty (u_i \,, u_1) , \qquad
\Delta \tilde u_Q = e^{- i p_{Q, \infty} L} \mathop {\rm Res} \limits_{u_1 = u_{1, \infty}} \prod_{i=1}^{Q-1} S_\infty (u_Q \,, u_i) ,
\label{multi-L Delta uw}
\end{equation}
where we used the unitarity relation $S (u, v)^{-1} = S (v, u)$. From \eqref{Q-bound rapidity} we have ${\rm Im} \, p_{1, \infty} > 0$ and ${\rm Im} \, p_{Q, \infty} < 0$. Thus we find $\Delta \tilde u_1$ and $\Delta \tilde u_Q$ are exponentially suppressed in $L$.

Let $\epsilon (p) \equiv \varepsilon (u (p))$ be the dispersion relation of an elementary magnon at $L=\infty$, and
\begin{equation}
\sum_{j=1}^{Q} \epsilon (p_j) = \epsilon_Q (P) \equiv \varepsilon_Q (U (P)),
\label{multi-L boundstate dispersion}
\end{equation}
be the dispersion relation of a $Q$-particle boundstate, which depends solely on $U = U(P)$. We also have
\begin{equation}
\sum_{j=1}^Q \(\Delta p_{j,U}\) \frac{d \epsilon (p_j)}{d p_j} =
\sum_{j=1}^Q \(\Delta U\) \frac{d \varepsilon (u_j)}{d u_j} =
\(\Delta U\) \frac{d}{d u_j} \, \sum_{j=1}^Q \varepsilon (u_j) =
\(\Delta P_U\) \frac{d \epsilon_Q (P)}{d P} \,.
\label{multi-L epsilonQ}
\end{equation}

\bigskip
The finite-size correction to the energy of the state $A$ is defined by
\begin{equation}
\Delta E_{\rm Bethe} = \sum_{j=1}^Q (\Delta p_j) \, \frac{d \epsilon (p_{j, \infty})}{d p_{j, \infty}} \,.
\label{energy Bethe}
\end{equation}
With the help of the relations \eqref{multi-L epsilonQ} and \eqref{multi-L Delta P}, one can show
\begin{align}
\Delta E^a &= (\Delta \tilde p_1) \, \frac{d \epsilon (p_{1, \infty})}{d p_{1, \infty}}
+ (\Delta \tilde p_Q) \, \frac{d \epsilon (p_{Q, \infty})}{d p_{Q, \infty}}
+ (\Delta P_U) \, \frac{d \epsilon_Q (P)}{d P}
\notag \\[2mm]
&=
(\Delta \tilde p_1^{\, a}) \( \frac{d \epsilon (p_{1, \infty}^a)}{d p_{1, \infty}^a} - \frac{d \epsilon_{Q_a} (P^a_\infty)}{d P^a_\infty}\) +
(\Delta \tilde p^{\, a}_Q) \( \frac{d \epsilon (p_{Q_a, \infty}^a)}{d p_{Q_a, \infty}^a} - \frac{d \epsilon_{Q_a} (P^a_\infty)}{d P^a_\infty} \)  \,.
\label{multi-L fs energy 1}
\end{align}
To obtain a simpler expression, let us make two more assumptions on the momenta and the $S$-matrix. Namely, $p_{Q + 1 - j} = (p_j)^*$ for the momenta of magnons constituting a boundstate, and
\begin{gather}
S (p_i \,, p_j)^* = \frac{1}{S ((p_i)^* \,, (p_j)^*)} = \frac{1}{S (p_{Q + 1 - i} \,, p_{Q + 1 - j})} \,,
\label{multi-L S-matrix reality}
\end{gather}
for the $S$-matrix. Under these assumptions one can show $\Delta \tilde p_Q = (\Delta \tilde p_1)^*$. However, one should keep it in mind that these assumptions do not hold in general. For instance, it was shown that a two-magnon boundstate at strong coupling do not generally obey $p_1 = (p_2)^*$ \cite{AF07}.

By substituting the result for $\Delta p_w = \(\Delta u_w \) p' (u_w) \ (w=1, Q)$ given in \eqref{multi-L Delta uw} together with $\Delta \tilde p_Q = (\Delta \tilde p_1)^*$, we finally arrive at the expression
\begin{equation}
\Delta E_{\rm Bethe} \approx 2 \ssp {\rm Re} \, \Bigg\{ \{ \epsilon' (p_{Q, \infty}) - \epsilon'_Q (P_\infty) \} \, e^{- i p_{Q, \infty} L} \, \mathop {\rm Res} \limits_{p_Q = p_{Q, \infty}} \prod_{k = 1}^{Q-1} S_\infty (p_Q \,, p_k) \Bigg\} .
\label{multi-L fs energy 2}
\end{equation}
This is the leading finite-size correction to the energy of a $Q$-particle boundstate, and is equivalent to the one obtained in \cite{MO08}. There they computed the finite-size correction for Heisenberg spin chain with $Q \gg g$, and found it agrees with the finite-size correction to a dyonic giant magnon.

\bigskip
For comparison, here we quote the generalized L\"uscher $\mu$-term for a $Q$-particle boundstate:
\begin{equation}
\delta E_A^\mu \approx \sum_b \sum_{\rm all\ residues} {\rm Re} \, \Bigg\{ (-1)^{F_b} \, \{ \epsilon_b' (q_*) - \epsilon'_Q (P) \} \, e^{- i q_* L} \, \mathop {\rm Res} \limits_{q = q_* (P)} S_{b A}^{\, b A} (q, P) \Bigg\} .
\label{multi-L mu energy}
\end{equation}
In \cite{HS08}, it was shown that the sum over the flavors of $b$ and the sum over the residues of $S$-matrix provide the factor of 2 at strong coupling, which implies an interesting observation $\delta E_A^\mu \approx \Delta E_{\rm Bethe}$\,.

\bigskip
This equality should be regarded as approximate and not rigorous, as we discussed at the beginning of this section. Consider, for instance, the case of $Q=1$. The result of the Bethe Ansatz \eqref{multi-L fs energy 2} is insensible because no $S$-matrix is present, while the $\mu$-term \eqref{multi-L mu energy} can still predict the finite-size correction to a single giant magnon.

\subsubsection{Multi-particle states}

We will illustrate how we can generalize the above result into the multi-particle states. Let $A = \{ A_1 (P^1) \ldots A_M (P^M) \}$ be the state composed of magnons and magnon boundstates where at least one of the $A_j$'s is a boundstate. Each $A_a (P^a)$ is a $Q_a$-particle boundstate, whose constituent magnons are written as $\{ p_1^a \,, \ldots \,, p_{Q_a}^a \}$. When the size of spin chain $L$ becomes finite, the momentum of the constituent magnons begins to fluctuate around their position at $L=\infty$ as
\begin{equation}
p_{j, L}^a = p_{j,\infty}^a + \Delta p_j^a \equiv p_{j,\infty}^a + \Delta \tilde p_j^a + \Delta p_{j,U}^a \,.
\end{equation}
Using the rapidity variables, this can be rewritten as
\begin{equation}
\Delta u^a_k = \Delta \tilde u_k^{\, a} + \Delta U^a \,,
\end{equation}
where $\Delta p^a_{k. U}$ accounts for the overall shift $\Delta U^a$. A new feature of the multi-particle states is that the momentum of each $Q_a$-particle boundstate may possibly fluctuate:
\begin{equation}
P_L^a = \sum_{j=1}^{Q_a} p_{j,L}^a \equiv P_\infty^a + \Delta P^a \,.
\end{equation}
Note the total momentum of the whole system is quantized as
\begin{equation}
P_{\rm total} = \sum_{a=1}^M P_L^a = \sum_{a=1}^M P_\infty^a = \frac{2 \pi I_{\rm total}}{L} \qquad \( I_{\rm total} \in \bb{Z} \).
\end{equation}

The Bethe Ansatz equations are simply written as
\begin{equation}
1 = e^{-i p_j^a L} \prod_{k \neq j}^{Q_a} S_L (p_j^a \,, p_k^a) \prod_{b \neq a}^M \prod_{\ell = 1}^{Q_b} S_L (p_j^a \,, p_\ell^b) .
\label{multi-L BAE 1}
\end{equation}
Evaluating the $S$-matrix in \eqref{multi-L BAE 1} around the boundstate pole, we can determine the variations $\Delta \tilde p_1^{\, a}$ and $\Delta \tilde p_{Q_a}^{\, a}$ as in \eqref{multi-L Delta uw}:
\begin{alignat}{3}
\Delta \tilde p_1^{\, a} &= \ e^{i p_{1, \infty}^a L} &\, &\mathop {\rm Res} \limits_{p_1^a = p_{1, \infty}^a} \pare{ \prod_{k = 2}^{Q_a} S_L (p_k^a \,, p_1^a) \prod_{b \neq a}^M \prod_{\ell = 1}^{Q_b} S_L (p_\ell^b \,, p_1^a) } ,
\label{multi-L Delta pwa1} \\[1mm]
\Delta \tilde p_{Q_a}^{\, a} &= e^{- i p_{Q_a, \infty}^a L} &\, &\mathop {\rm Res} \limits_{p_{Q_a}^a = \, p_{Q_a, \infty}^a} \pare{ \prod_{k=1}^{Q_a-1} S_L (p_{Q_a}^a \,, p_k^a) \prod_{b \neq a}^M \prod_{\ell = 1}^{Q_b} S_L (p_{Q_a}^a \,, p_\ell^b) } ,
\label{multi-L Delta pwaQ}
\end{alignat}
Let us consider the product of \eqref{multi-L BAE 1} over $j=1,\ldots , Q_a$ and take its logarithm. It gives
\begin{equation}
- 2 \pi I^a =  P_L^a L - \sum_{j=1}^{Q_a} \sum_{b \neq a}^M \sum_{\ell = 1}^{Q_b} \delta_L (p_j^a \,, p_\ell^b) \,,
\label{multi-L BAE 3}
\end{equation}
where $I^a$ is an integer and $i\delta_L (p, q) \equiv \ln S_L (p, q)$. Considering an infinitesimal variation of this equation, we immediately find
\begin{equation}
0 = (\Delta P^a) L + P^a (\Delta L) - \sum_{j=1}^{Q_a} \sum_{b \neq a}^M \sum_{\ell = 1}^{Q_b} \pare{ (\Delta p_j^a) \, \frac{\partial \delta_L (p_j^a \,, p_\ell^b)}{\partial p_j^a} + (\Delta p_\ell^b) \, \frac{\partial \delta_L (p_j^a \,, p_\ell^b)}{\partial p_\ell^b} },
\end{equation}
which shows $\Delta P^a \approx \Delta p_j^a/L$ or $\Delta L/L$, which is negligible for large $L$. Hence, we can set $\Delta P^a \approx 0$ and treat the finite-size effects for each boundstate of $A$ separately as long as $L$ is large. As a by-product of this argument, one can see that the finite-size effects for elementary particles, namely $\Delta P^b$ for $Q_b = 1$, are negligible for large $L$.

The finite-size correction to the total energy is
\begin{equation}
\Delta E_{\rm Bethe} = \sum_{a=1}^M \Delta E_{A_a} = \sum_{a=1}^M \sum_{j=1}^{Q_a} \(\Delta p_j^a\) \frac{d \epsilon (p_{j, \infty}^a)}{d p_{j, \infty}^a} \,,
\label{multi-L energy Bethe}
\end{equation}
and each of $\Delta E_{A_a}$'s can be evaluated in much the same way as in one-particle states, \eqref{multi-L fs energy 1} or \eqref{multi-L fs energy 2}. The major difference from one-particle states is that the displacements of the momentum \eqref{multi-L Delta pwa1}, \eqref{multi-L Delta pwaQ} acquire a lot more $S$-matrix factors in their right hand sides. Assuming again the reality conditions on the momentum and the $S$-matrix \eqref{multi-L S-matrix reality}, the finite-size correction \eqref{multi-L energy Bethe} becomes
\begin{multline}
\Delta E_{\rm Bethe} \approx 2 \ssp {\rm Re} \, \Bigg\{ \sum_{a=1}^M \pare{ \epsilon' (p_{Q_a, \infty}^{\,a}) - \epsilon'_{Q_a} (P_\infty^a) } e^{- i p_{Q_a, \infty}^{\,a} L} \; \times \\
\mathop {\rm Res} \limits_{p_{Q_a}^{\,a} = p_{Q_a, \infty}^{\,a}} \prod_{k = 1}^{Q_a-1} S_\infty (p_{Q_a}^{\,a} \,, p_k^{\,a}) 
\prod_{b \neq a}^M \prod_{\ell = 1}^{Q_b} S_L (p_{Q_a}^{\,a} \,, p_\ell^{\,b} ) \Bigg\} .
\label{multi-L energy Bethe 2}
\end{multline}

Just like one-particle states, one can repeat the arguments that led to $\delta E_A^\mu \approx \Delta E_{\rm Bethe}$\,, to replace the factor of 2 by the sums over flavors and residues.
Combining this argument as well as the results in Section \ref{sec:classical mgm}, we conjecture the $\mu$-term formula for multi-magnon states shall be given by \eqref{multi mu-term1}, namely
\begin{equation}
\delta E_A^\mu \approx {\rm Re} \pare{ \sum_{\ell=1}^M \sum_b (-1)^{F_b} \pare{\epsilon_b' (q_\ell^*) - \epsilon'_{a_\ell} (p_\ell)} e^{- i q_\ell^* L} \mathop {\rm Res} \limits_{q^1 = q_\ell^*} S_{b a_\ell}^{\, b a_\ell} (q^1 \,, p_\ell) \prod_{k \neq \ell} S_{b a_k}^{\, b a_k} (q_\ell^* \,, p_k) },
\label{multi mu-term3}
\end{equation}
where the sum over all possible residues are understood implicitly.

\section{Summary and Discussions}\label{sec:summary}

In this paper, we proposed the generalized L\"uscher formula for multi-particle states. The formula consists of $F$-term and $\mu$-term, which correspond to one-loop and classical finite-size correction to the energy of a string state, respectively.

In Section \ref{sec:F-term}, we followed the finite-gap method of \cite{GSV08} to obtain the $F$-term for multi-particle states. The $F$-term formula was then compared with the exact finite-size spectrum of the sinh-Gordon theory \cite{Teschner07}, and the agreement is found.
In Section \ref{sec:mu-term}, in search of the correct $\mu$-term formula we considered the finite-size correction of multi (dyonic) giant magnons computed in \cite{MO08}, and calculated the finite-size correction to the energy of the states with many magnon boundstates by using the asymptotic Bethe Ansatz in the $su(2)$ sector. It was shown that our $\mu$-term formula for multi-particle states is consistent with both results.
Also, in Appendix \ref{sec:wrapping}, we shall show that various transcendental terms which appeared as the wrapping effects for the length four Konishi operator of $\cN=4$ super Yang-Mills theory, can partially be reproduced by evaluating the $F$-term formula at weak coupling.

As we argued in the introduction, these formulae will be regarded as the large size limit of the TBA equations for the $su(2|2)^2$ spin chain. It will be important to give rigorous derivation of these formulae in order to compute finite-size effects more precisely.

Finite-size effects in general will give further insight into the AdS/CFT correspondence. Recall that the BPS condition imposes strict constraints on the dispersion relation at infinite size \eqref{magnon dispersion:gauge} and \eqref{magnon dispersion:string}. The finite-size corrections, in contrast, contain dynamical information of the theory. For instance, once the finite-size spectrum is obtained, we may use the formula in the opposite direction, to probe the spectrum of the (mirror) theory, or the on-shell splitting processes.

Since the finite-size effects depend sensitively on boundary conditions, it is very interesting to apply the generalized L\"uscher formula to integrable, exactly marginal deformations of the $\cN=4$ super Yang-Mills theory. One famous example is the $\beta$ (or the Leigh-Strassler) deformation \cite{LS95}, and various directions of integrable, exactly marginal deformation are known \cite{Roiban03, BC04, BR05a, FKM05, BM05, MS06, Mansson07a, Mansson07b}. The supergravity dual is obtained by a sequence of $T$-dualities \cite{LM05}, and the integrability of classical string action is studied in \cite{FRT05a, Frolov05, AAF05}. Notably, one can write down the $S$-matrix in the same way as $\cN=4$ while boundary condition is twisted. Moreover, wrapping effects \cite{FSSZ08b} as well as finite-size effects \cite{BF08} have already been known for certain deformations, which should be reproduced from the L\"uscher-type computation.

\subsubsection*{Acknowledgments}

We would like to acknowledge Stefano Kovacs and Sergey Frolov for carefully reading the draft. R.S. is grateful for Sergey Frolov for interesting discussions.
The work of Y.H. is supported in part by JSPS Research Fellowships for Young Scientists. The work of R.S. is supported by the Science Foundation Ireland under Grant No.07/RFP/PHYF104.

\appendix

\section{Kinematical factor for multi giant magnon background}\label{app:FG}

Following the same step and the same notation as \cite{GSV08}, we can constrain the fluctuation of quasi-momenta as
\begin{align*}
\delta p_{\tilde 1} &= +\frac{A x +B }{x^2-1}-\sum_{n,j=\tilde 3\tilde 4\hat 3\hat 4} \(\frac{N_n^{\tilde 1 j} \alpha(x_n^{\tilde 1j})}{x-x_n^{\tilde 1 j}}-\frac{N_n^{\tilde 2 j} \alpha(x_n^{\tilde 2j})}{1/x-x_n^{\tilde 2 j}}-\frac{N_n^{\tilde 2 j} \alpha(x_n^{\tilde 2 j})}{x_n^{\tilde 2 j}}\) + \delta p_{\tilde 1}^{\rm GM} \,, \\
\delta p_{\tilde 2} &= +\frac{A x +B}{x^2-1}-  \sum_{n,j=\tilde 3\tilde 4\hat 3\hat 4} \(\frac{N_n^{\tilde 2 j}
\alpha(x_n^{\tilde 2j})}{x-x_n^{\tilde 2 j}}-\frac{N_n^{\tilde 1 j}
\alpha(x_n^{\tilde 1j})}{1/x-x_n^{\tilde 1 j}}-\frac{N_n^{\tilde 1 j} \alpha(x_n^{\tilde 1 j})}{x_n^{\tilde 1 j}}\) + \delta p_{\tilde 2}^{\rm GM} \,, \\
\delta p_{\tilde 3} &= -\frac{C x +D}{x^2-1}+\sum_{n,j=\tilde 1\tilde 2\hat 1\hat 2} \(\frac{N_n^{\tilde 3 j} \alpha(x_n^{\tilde 3j})}{x-x_n^{\tilde 3 j}}-\frac{N_n^{\tilde 4 j} \alpha(x_n^{\tilde 4j})}{1/x-x_n^{\tilde 4 j}}-\frac{N_n^{\tilde 4 j} \alpha(x_n^{\tilde 4 j})}{x_n^{\tilde 4 j}}\) + \delta p_{\tilde 3}^{\rm GM} \,, \\
\delta p_{\tilde 4} &= -\frac{C x +D }{x^2-1}+\sum_{n,j=\tilde 1\tilde 2\hat 1\hat 2} \(\frac{N_n^{\tilde 4 j} \alpha(x_n^{\tilde 4 j})}{x-x_n^{\tilde 4 j}}-\frac{N_n^{\tilde 3 j} \alpha(x_n^{\tilde 3 j})}{1/x-x_n^{\tilde 3 j}}-\frac{N_n^{\tilde 3 j} \alpha(x_n^{\tilde 3 j})}{x_n^{\tilde 3 j}}\) + \delta p_{\tilde 4}^{\rm GM} \,, \\
\delta p_{\hat 1} &= +\frac{A x +B }{x^2-1}+ \sum_{n,j=\hat 3\hat 3\tilde 3 \tilde 4} \(\frac{N_n^{\hat 1 j} \alpha(x_n^{\hat 1 j})}{x-x_n^{\hat 1 j}}-\frac{N_n^{\hat 2 j} \alpha(x_n^{\hat 2 j})}{1/x-x_n^{\hat 2 j}}-\frac{N_n^{\hat 2 j} \alpha(x_n^{\hat 2j})}{x_n^{\hat 2 j}}\) , \\
\delta p_{\hat 2} &= +\frac{A x +B}{x^2-1} + \sum_{n, j=\hat 3\hat 4\tilde 3\tilde 4} \(\frac{N_n^{\hat 2 j} \alpha(x_n^{\hat 2 j})}{x-x_n^{\hat 2 j}}-\frac{N_n^{\hat 1 j} \alpha(x_n^{\hat 1 j})}{1/x-x_n^{\hat 1 j}}-\frac{N_n^{\hat 1 j} \alpha(x_n^{\hat 1j})}{x_n^{\hat 1 j}}\) , \\
\delta p_{\hat 3} &= -\frac{C x +D}{x^2-1}- \sum_{n,j=\hat 1\hat 2\tilde 1\tilde 2} \(\frac{N_n^{\hat 3 j} \alpha(x_n^{\hat 3 j})}{x-x_n^{\hat 3 j}}-\frac{N_n^{\hat 4 j} \alpha(x_n^{\hat 4 j})}{1/x-x_n^{\hat 4 j}}-\frac{N_n^{\hat 4 j} \alpha(x_n^{\hat 4 j})}{x_n^{\hat 4 j}}\) , \\
\delta p_{\hat 4} &= -\frac{C x +D}{x^2-1}-  \sum_{n,j=\hat 1\hat 2\tilde 1\tilde 2} \(\frac{N_n^{\hat 4 j} \alpha(x_n^{\hat 4j})}{x-x_n^{\hat 4 j}}-\frac{N_n^{\hat 3 j} \alpha(x_n^{\hat 3 j})}{1/x-x_n^{\hat 3 j}}-\frac{N_n^{\hat 3 j} \alpha(x_n^{\hat 3 j})}{x_n^{\hat 3 j}}\) ,
\end{align*}
where $\delta p_i^{\rm GM}$ account for the backreaction to classical background of multi giant magnons. The $\delta p_i^{\rm GM}$'s are given by
\begin{alignat*}{2}
\delta p_{\tilde 1}^{\rm GM} &= - \sum_{\ell=1}^N \sum_{\beta=\pm } \(\frac{A_\ell^\beta}{1/x-X_\ell^\beta} + \frac{A_\ell^\beta}{X_\ell^\beta} \) , &\qquad
\delta p_{\tilde 2}^{\rm GM} &= \sum_{\ell=1}^N \sum_{\beta=\pm } \frac{A_\ell^\beta}{x-X_\ell^\beta} \,, \\
\delta p_{\tilde 3}^{\rm GM} &= - \sum_{\ell=1}^N \sum_{\beta=\pm } \frac{A_\ell^\beta}{x-X_\ell^\beta} \,, &\qquad
\delta p_{\tilde 4}^{\rm GM} &= \sum_{\ell=1}^N \sum_{\beta=\pm } \(\frac{A_\ell^\beta}{1/x-X_\ell^\beta} + \frac{A_\ell^\beta}{X_\ell^\beta}\) .
\end{alignat*}

As discussed in \cite{GV07b, GSV08}, the parameters $A,B,C,D$ and $A_\ell^\beta$ are constrained from $x \to 1/x$ symmetry and the large $x$ asymptotics. Concerning $A_\ell^\beta$, we obtain the relations
\begin{align}
\cN_{\rm all} &\equiv \sum_n \sum_{i=\tilde 1 \tilde 2 \hat 1 \hat 2} \sum_{j=\tilde 3 \tilde 4 \hat 3 \hat 4} N_n^{ij} \, \frac{ \alpha(x_n^{ij}) }{x_n^{ij}} = \sum_{\ell=1}^M \sum_{\beta=\pm} \frac{A_\ell^\beta}{X_\ell^\beta} \,,
\label{Alb rel1} \\
0 &= \sum_{\ell=1}^M \sum_{\beta=\pm} A_\ell^\beta \( 1 - \frac{1}{(X_\ell^\beta)^2} \).
\label{Alb rel2}
\end{align}
These relations can be solved as
\begin{equation}
\frac{A_\ell^+}{\cN_{\rm all}} = \alpha_\ell \, \frac{(X_\ell^+)^2 (X_\ell^- - 1)(X_\ell^- + 1)}{(X_\ell^- - X_\ell^+)(X_\ell^- X_\ell^+ + 1)} \,, \quad
\frac{A_\ell^-}{\cN_{\rm all}} = - \alpha_\ell \, \frac{(X_\ell^-)^2 (X_\ell^+ - 1)(X_\ell^+ + 1)}{(X_\ell^- - X_\ell^+)(X_\ell^- X_\ell^+ +1)} \,, \quad
\sum_{\ell=1}^M \alpha_\ell = 1 \,.
\end{equation}
Note, however, that we cannot determine each $\alpha_\ell$ only from the conditions discussed above.

The one-loop energy is expressed as
\begin{equation}
\delta \Delta = 2g \(\sum_n \sum_{i=\tilde 1 \tilde 2 \hat 1 \hat 2} \sum_{j=\tilde 3 \tilde 4 \hat 3 \hat 4} N_n^{ij} \, \frac{ \alpha(x_n^{ij}) }{(x_n^{ij})^2} - \sum_{\ell=1}^M \sum_{\beta=\pm} A_\ell^\beta \)
\equiv \sum_{i, j} \sum_n N_n^{ij} \, \Omega (x_n^{ij}) \,,
\label{dDelta in Omega}
\end{equation}
where
\begin{equation}
\Omega (x) = \frac{2}{x^2-1} \[ 1 - \sum_{\ell=1}^M \alpha_\ell \(\frac{X_\ell^- + X_\ell^+}{X_\ell^- X_\ell^+ +1}\) x \].
\label{M-mag Omega:app}
\end{equation}

\section{Notation for the $su(2)$ Bethe Ansatz}\label{sec:notation}

Our notation is similar to the one used in \cite{BS05}.

Let us introduce the rapidity variable $u(p)$ by
\begin{equation}
u (p) = \frac{1}{2} \, \cot \frac{p}{2} \sqrt{1 + 16 g^2 \sin^2 \frac{p}{2}} \qquad g \equiv \frac{\sqrt \lambda}{4 \pi} \,.
\label{def:rapidity up}
\end{equation}
Its Zhukovsky map of $x(u)$ and the spectral parameters $x^\pm$ are defined by
\begin{align}
u &= g \( x + \frac{1}{x} \) = \frac{g}{2} \( x^+ + \frac{1}{x^+} + x^- + \frac{1}{x^-} \) , 
\label{def:rapidity u} \\[1mm]
x^\pm &= x \(u = u(p) \pm \frac{i}{2} \) = e^{\pm ip/2} \, \frac{1 + \sqrt{1 + 16 g^2 \sin^2 \frac{p}{2}}}{4 g \sin \frac{p}{2}} \,.
\end{align}
The spectral parameters satisfy the identity
\begin{equation}
x^+ + \frac{1}{x^+} - x^- - \frac{1}{x^-} = \frac{i}{g} \,.
\label{spec identity}
\end{equation}

Consider a set of the spectral parameters $\{ x_1 \,, \ldots \,, x_Q \}$ which satisfy the boundstate condition $x_j^+ = x_{j+1}^-$\,. If we denote the outermost parameters by $X^+ = x_Q^+$ and $X^- = x_1^-$\,, then they take the form
\begin{equation}
X^\pm = e^{\pm iP/2} \, \frac{Q + \sqrt{Q^2 + 16 g^2 \sin^2 \frac{P}{2}}}{4 g \sin \frac{P}{2}} \qquad P = \sum_{k=1}^Q p_k \,,
\end{equation}
which can be shown using \eqref{spec identity}. The rapidity variable $U(P)$ for a boundstate is
\begin{equation}
U = \sum_{k=1}^Q u_k = \frac{g}{2} \( X^+ + \frac{1}{X^+} + X^- + \frac{1}{X^-} \) .
\label{def:rapidity U}
\end{equation}
The rapidities $\{ u_k \}$ which constitute a $Q$-particle boundstate can be written as
\begin{equation}
u_k = U + i \( k - \frac{Q+1}{2} \),
\end{equation}
which can be shown from the conditions $x_j^+ = x_{j+1}^-$ and the identities
\begin{equation}
u_1 - u_2 \pm i = \( x_1^\pm - x_2^\mp \) \( 1 - \frac{1}{x_1^\pm x_2^\mp} \) .
\end{equation}

\section{Wrapping and Transcendentality}\label{sec:wrapping}

As we saw in Section \ref{sec:luscher}, the generalized L\"uscher formula at weak coupling predicts the corrections of order $g^{2L}$ to the prediction of the asymptotic Bethe Ansatz, which are thought of as wrapping effects \cite{AJK05}.

In this appendix, we apply the generalized L\"{u}scher formula to length $L$ operators in the $su(2)$ sector at weak coupling, under several approximations.
Although our computation is not quantitatively rigorous, we can reproduce the transcendental terms that appear only after wrapping effects are taken into account \cite{FSSZ07, FSSZ08a, KM08, FSSZ08b}.

\subsection{From TBA to the L\"uscher formula}

The generalized L\"uscher formulae presented in Section \ref{sec:luscher} contain a sum over the virtual particle $b$. We were able to neglect terms with $Q_b > 1$ at strong coupling because they decay more rapidly than those with $Q_b = 1$. However, as we saw in \eqref{exp suppression}, at weak coupling they bring another contribution of order $g^{2L}$. Thus, we have to reconsider if we should include virtual particles with higher multiplet numbers.

In the perturbative computation in quantum field theory, of course, virtual particles are elementary fields which appear in the Lagrangian. As for integrable systems which are not defined in terms of Lagrangian, it is not clear if we may neglect virtual boundstate particles.
To answer this question, we recall that the (generalized) L\"uscher formula is regarded as the large size limit of TBA for excited states.

The starting point of TBA is the observation that the (Euclidean) partition function of a two dimensional theory is invariant under the interchange of the spacetime coordinates in two dimensions. One can thus equate the groundstate energy of the original theory at finite size with the free energy of the interchanged theory (also called the `mirror' theory \cite{AF07}) at finite temperature \cite{Zamolodchikov90}. Clearly, to compute a partition function or a free energy we need the whole spectrum of the theory including (stable) boundstates. The partition function obtained in this way has a sum over the particle spectrum and an integral over the momentum of the particle. Moreover, it is known that TBA for excited states is obtained by deforming the momentum integral of this partition function to pick up a pole singularity corresponding to a physical particle of the theory \cite{DT96, DT97}. Thus, the sum over spectrum should remain as before in the TBA formula for one-particle or multi-particle states.

Thus we conjecture that a sum over an infinite tower of BPS boundstates is necessary for the generalized L\"uscher formula to compute the wrapping effects correctly. Below we will see that this summation indeed reproduces the expected transcendental structure \eqref{wrap-trans:quote}.

\bigskip
Here is another remark on the mirror theory.
Upon identifying the generalized L\"uscher formula as a limiting behavior of TBA equations, we should reinterpret the spectrum, the dispersion relation and the $S$-matrix appearing in the generalized $F$-term formula as those of the mirror theory. On the other hand, we have used the $S$-matrix of the original $su(2|2)^2$ theory to compute the semiclassical spectrum of finite-size giant magnons.
Putting these two facts together, the conjecture of \cite{AF07} turns out to be very plausible; the mirror $S$-matrix is related to the original $S$-matrix via analytic continuation.

\subsection{The $\bmt{F}$-term at weak coupling and transcendentality}

We will focus on length $L\;(L=4,5,6,\dots)$ operators with two impurities with real momenta which have the form
\begin{equation}
{\rm tr} [W W Z^{J_1}] + \({\rm permutations}\), \quad L = J_1 + 2\,,
\label{two-magnon su2 states}
\end{equation}
where $W$ and $Z$ are complex scalars of ${\cN=4}$ super Yang-Mills. We then study how the transcendental wrapping effects appear from the L\"uscher formula.

First of all, we argue that there is no contribution from the $\mu$ term for the operators \eqref{two-magnon su2 states} at weak coupling. Recall that this operator can be interpreted as the state with two magnons of real momenta, $\{ a_1 (p_1) a_2 (p_2) \}$. Because the $\mu$-term for multi-magnon states is associated with the splitting process of either $a_1 (p_1)$ or $a_2 (p_2)$, it is sufficient to examine if the splitting of an elementary magnon occurs at weak coupling.

Let us look at the energy-momentum conservation at the point of splitting:
\begin{equation}
\sqrt{1+16g^2 \sin^2 \left( \frac{p_a}{2} \right)}=\sqrt{Q_b^2+16g^2 \sin^2 \left( \frac{p_b}{2} \right)}
+\sqrt{Q_c^2+16g^2 \sin^2 \left( \frac{p_a-p_b}{2} \right)}\,.
\label{eq:on-shell}
\end{equation}
This equation can be solved by $x_b^\pm=1/x_a^\mp$, and one of these conditions correspond to a pole of $S$-matrix. However, this is not a physical pole in the sense that at least a pair of spectral parameters are found in the unphysical region $\abs{x_j^\pm} < 1$ \cite{AF07}.
We can conclude that the $\mu$-term does not contribute at weak coupling.

\bigskip
Thus, what matters is calculation of the $F$-term. In this paper, our analysis is restricted to the case where the virtual particle is one of the symmetric scalar components in the boundstate multiplet, where one can use the fusion rule to obtain elementary-boundstate $S$-matrix. Of course, we have to sum over all $16 Q_b^2$ polarizations of the (mirror) $Q_b$\,-boundstate in order to obtain the quantitatively correct answer. It is nevertheless remarkable that we are able to capture the appearance of transcendentality only from such a simple computation.
Under this assumption,\footnote{
Strictly speaking, our assumptions include: (i) an infinite tower of BPS boundstates completes the spectrum of the $su(2|2)$ spin chain. (ii) the spectrum of the mirror theory is same as the original theory. (iii) the dispersion relation of the mirror particle is given by the Wick rotation} the $F$-term can be rewritten as
\begin{multline}
\delta E^F_{su(2)} \approx - \sum_{Q_b=1}^\infty (Q_b + 1)^2 \ \times \\
\int_{-\infty}^\infty \frac{d \tilde q}{2\pi} \left( 1-\sum_{k=1}^M \alpha_k \, \frac{\epsilon_{a_k}' (p_k)}{\epsilon_b' (q^1)} \right)
\, e^{-2L\arcsinh \Bigl(\frac{\sqrt{Q_b^2 + \tilde q^2}}{4g}\Bigr)}
\prod_{\ell=1}^2 S_{ba_\ell}^{ba_\ell} (q^1,p_\ell) \,.
\end{multline}
where factor $(Q_b+1)^2$ comes from the degeneracy of symmetric scalars in $Q_b$\,-boundstate, $ \phi^{( i_1} \ldots \phi^{i_{Q_b} )} \, \bar \phi^{\, ( \bar \jmath_1} \ldots \bar \phi^{\, \bar \jmath_{Q_b})}$ with $i_a \,, \bar \jmath_b = 1$ or $2$.

At leading order of $g \ll 1$, the exponential factor becomes
\begin{align}
e^{-2L\arcsinh \Bigl(\frac{\sqrt{Q_b^2+ \tilde q^2}}{4g}\Bigr)} \simeq \frac{(4 g^2)^L}{(Q_b^2+ \tilde q^2)^L}\,.
\label{main factor}
\end{align}
and the backreaction terms become
\begin{equation}
\sum_{k=1}^2 \alpha_k \, \frac{\epsilon_{a_k}' (p_k)}{\epsilon_b' (q^1)} \, e^{-2L\arcsinh \Bigl(\frac{\sqrt{Q_b^2 + \tilde q^2}}{4g}\Bigr)} \approx - \sum_{k=1}^2 \alpha_k \cdot 2 \tilde q \sin (p_k) \, \frac{(4 g^2)^{L+1}}{(Q_b^2+ \tilde q^2)^{L+1}} \,.
\label{backreaction factor}
\end{equation}
Assuming $\alpha_k$ is regular at $g=0$, the backreaction part is higher order in $g$ compared to the main part \eqref{main factor}. So we neglect the backreaction part in what follows.

The elementary-boundstate $S$-matrix is same as the one in the $su(2)$ sector, and is given by%
\footnote{Note added: It was pointed out in \cite{BJ08} that we should use $S$-matrix in the $sl(2)$ sector, because boundstates in the mirror model live in ${\rm AdS}_5$ subspace \cite{AF07}. Results of our primitive computation do not change even if we use the $S$-matrix in the $sl(2)$ sector.}
\begin{equation}
S_{ba_l}^{ba_l}(q^1,p_l)=G_l(Q_b-1) G_l(Q_b+1)\sigma^2(Y_b,X_{a_l}),
\end{equation}
where
\begin{align}
G_l(Q)=\frac{u(q^1,Q_b)-u(p_l,1)+iQ}{u(q^1,Q_b)-u(p_l,1)-iQ}
\;\;\;\; {\rm with} \;\;\;\;
u(p,Q)=\frac{1}{2}\cot \left( \frac{p}{2} \right) \sqrt{Q^2+16g^2 \sin^2\left( \frac{p}{2} \right)} \,,
\end{align}
and $\sigma^2(Y_b,X_{a_l})$ is the dressing phase, which does not contribute to the $F$-term at $g^{2L}$. Then using $u(q^1,Q_b) \approx - \tilde q/2$, the $S$-matrix factor is evaluated as
\begin{equation}
\prod_{\ell=1}^2 S_{ba_\ell}^{ba_\ell}(q^1,p_\ell) = \prod_{\ell=1}^2 \frac{- \tilde q/2 -u(p_l,1)+i(Q_b-1)}{- \tilde q/2 -u(p_l,1)-i(Q_b-1)} \cdot \frac{- \tilde q/2 -u(p_l,1)+i(Q_b+1)}{- \tilde q/2 -u(p_l,1)-i(Q_b+1)} \,,
\end{equation}
which tends to $1$ for $Q_b \gg 1$.

Thus after performing the integration over $\tilde{q}$, we obtain the final expression
\begin{align}
\delta E^F_{su(2)} &= - \sum_{Q_b=1}^\infty Q_b^2 \cdot \frac{1}{2\pi} \cdot (4 g^2)^L \cdot \frac{\sqrt{\pi}}{Q_b^{2L-1}} \frac{\Gamma(L-\frac{1}{2})}{\Gamma(L)} + \( \frac{1}{Q_b^{2L-2}} \) \\
&\approx - \frac{2^{2L-1} }{\sqrt{\pi}} \frac{\Gamma(L-\frac{1}{2})}{\Gamma(L)} \, \zeta (2L-3) \, g^{2L}\,.
\end{align}
If we substitute the values $L=4,5,6,7$ to this result, they become:
\begin{align*}
\delta E^F_{su(2)}(L=4)&\approx - 40 \, \zeta(5) \, g^8,\;\;\;\;\;
\delta E^F_{su(2)}(L=5) \approx - 140 \, \zeta(7) \, g^{10},\; \\
\delta E^F_{su(2)}(L=6)&\approx -  504 \, \zeta(9) \, g^{12},\;\;
\delta E^F_{su(2)}(L=7) \approx - 1848 \, \zeta(11) \, g^{14}\,.
\end{align*}
The results show that the $F$-term for length $L$ operator contains a term proportional to $\zeta (2L-3)$ as conjectured in \eqref{wrap-trans:quote}. There may possibly be additional terms of other transcendental degree, {\it e.g.} $\zeta(2L-2), \zeta(2L-1), \ldots$ or $\zeta(2L-4), \zeta(2L-5), \ldots$, if we include the contributions from the whole $S$-matrix and compute the $F$-term without any approximations.

\bigskip
From the standpoint of the AdS/CFT correspondence, it may be puzzling to substitute $L=J_1$ at strong coupling and $L=J_1+J_2$ at weak coupling. As discussed in \cite{HS08}, this is a consequence of the fact that finite-size effects depend on the choice of frame. In fact, we have seen a similar phenomenon also in Section \ref{sec:F-term}. There we found the choice of frame is related to that of twists. Different choice of twists should modify the finite-size effects, because they are physical quantities sensitive to boundary conditions.

\end{document}